%% file: ANA-EXOT-2019-04-PAPER.tex
\pdfoutput=1
\documentclass[PAPER, atlasdraft=false, texlive=2023, UKenglish, cernpreprint, texmf, orcidlogo]{atlasdoc}
 
\usepackage{atlaspackage}
\usepackage{atlasbiblatex}
 
\usepackage{atlasphysics}
 
\graphicspath{{logos/}{figures/}}
 
\usepackage{ANA-EXOT-2019-04-PAPER-defs}
 
\usepackage{diagbox}
\usepackage{float}
\usepackage{multirow}
 
\usepackage{amsmath,bm}
\usepackage{array}
\newcolumntype{C}[1]{>{\centering\let\newline\\\arraybackslash\hspace{0pt}}m{#1}}

\newcommand{\rcorr}{R_\text{corr}}
\newcommand\Tstrut{\rule{0pt}{2.6ex}}       
\newcommand\Bstrut{\rule[-0.9ex]{0pt}{0pt}} 
\newcommand{\TBstrut}{\Tstrut\Bstrut}

\AtlasJournal{JHEP}
\AtlasTitle{Search for single vector-like $B$ quark production and decay via $B\rightarrow bH(b\bar{b})$ in $pp$ collisions \\
at $\sqrt{s} = 13$~TeV with the ATLAS detector}

\AtlasJournalRef{J. High Energ. Phys. 2023, 168 (2023)}
\AtlasDOI{10.1007/JHEP11(2023)168}
 
\AtlasAbstract{
A search is presented for single production of a  vector-like $B$ quark decaying into a Standard Model $b$-quark and a Standard Model Higgs boson, which decays into a $b\bar{b}$ pair. The search is carried out in 139~fb$^{-1}$ of $\sqrt{s} = 13$~TeV proton--proton collision data collected by the ATLAS detector at the LHC between 2015 and 2018. No significant deviation from the Standard Model background prediction is observed, and mass-dependent exclusion limits at the $95\%$ confidence level are set on the resonance production cross-section in several theoretical scenarios determined by the couplings $c_W$, $c_Z$ and $c_H$ between the $B$ quark and the Standard Model $W$, $Z$ and Higgs bosons, respectively. For a vector-like $B$ occurring as an isospin singlet, the search excludes values of $c_W$ greater than 0.45 for a $B$ resonance mass ($m_B$) between 1.0 and 1.2~TeV. For $1.2~\text{TeV} < m_B < 2.0~\text{TeV}$, $c_W$ values larger than 0.50--0.65 are excluded. If the $B$ occurs as part of a $(B,Y)$ doublet,  the smallest excluded $c_Z$ coupling values range between 0.3 and 0.5 across the investigated resonance mass range $1.0~\text{TeV} < m_B < 2.0~\text{TeV}$.
}
 
\author{The ATLAS Collaboration}
 
\AtlasRefCode{EXOT-2019-04}
\PreprintIdNumber{CERN-EP-2023-091}

\hypersetup{pdftitle={ATLAS document},pdfauthor={The ATLAS Collaboration}}

\begin{document}
 
\title{Search for single vector-like $B$ quark production and decay via $B\rightarrow bH(b\bar{b})$ in $pp$ collisions \\
at $\sqrt{s} = 13$\,TeV with the ATLAS detector}
 
\maketitle
 
\tableofcontents
\newpage
 
\section{Introduction}
\label{sec:intro}

The observation of a particle compatible with the Higgs boson at the Large Hadron Collider (LHC)~\cite{HIGG-2012-27, CMS-HIG-12-028} completed the set of fundamental particles predicted to exist according to the Standard Model (SM). Nevertheless, the comparatively low observed Higgs boson mass conflicts with the expected effect of higher-order quantum-loop mass corrections, which would push the physical Higgs boson mass towards the Planck scale. Such observations and naturalness arguments~\cite{Susskind:1978ms} suggest the existence of an as-yet undiscovered mechanism beyond the SM preventing such divergent mass contributions~\cite{deGouvea:2014xba}. Theoretical extensions of the SM attempt to provide a natural solution to this issue by postulating the Higgs boson to be either a composite particle~\cite{Dugan:1984hq, Agashe:2004rs} or a pseudo Nambu--Goldstone boson, such as in the Little Higgs model~\cite{Schmaltz:2005ky}.
In these models, an additional symmetry corresponds to a new strong interaction, whose bound states include vector-like quarks (VLQs). 
 
VLQs are predicted to be spin-1/2 particles that transform as a triplet (hence \enquote{vector-like}) under colour gauge symmetry and whose left- and right-handed components both have the same electroweak quantum numbers~\cite{Aguilar-Saavedra:2013qpa}. They couple to the SM  fermions via Yukawa couplings~\cite{Cacciapaglia:2011fx} and therefore interact principally with the third-generation SM quarks. 
Theoretical constraints on the renormalisability of the coupling constants restrict the occurrence of VLQs to seven gauge-covariant multiplets under the weak hypercharge gauge symmetry. Vector-like $T$ and $B$ quarks, the vector-like equivalents of the third-generation SM quarks with electric charge $Q_T = 2/3$ and $Q_B = -1/3$, can exist as singlets, doublets or triplets, whereas the $X$ and $Y$ VLQs, with exotic charges $Q_X = 5/3$ and $Q_Y = -4/3$ respectively, can exist either in gauge doublets along with a $T$ or $B$ quark or in gauge triplets along with both the $T$ and $B$ quarks.
 
At the LHC, VLQs are expected to be produced either in pairs, via the strong interaction, or singly, via the exchange of an intermediate electroweak gauge boson. VLQ pair production is a pure QCD process with a cross-section that, at leading order, depends solely on the VLQ mass, whereas single-VLQ production cross-sections are also strongly affected by both the coupling strength to the SM quarks and the multiplet considered~\cite{Aguilar-Saavedra:2013qpa}, allowing the various theoretical scenarios to be probed in more detail. Furthermore, single VLQ production may overtake pair production as the principal VLQ production mechanism above a \TeV-mass threshold depending on the strengths of couplings to the SM quarks.
 
The theoretical framework for VLQs sets a common bare-mass term in the Yukawa Lagrangian for all VLQs, resulting in a mass splitting of the order of 1--10~\GeV\ among the various weak eigenstates. The main consequence of this feature is the heavy suppression of all cascade decays of one VLQ into another, which results in the total decay width being mainly determined by the coupling to the SM third-generation quarks.
For $T$ and $B$ VLQs, the allowed decays are neutral-current conversion into the SM equivalent ($T\rightarrow tH, tZ$ and $B\rightarrow bH, bZ$) or charged-current decays via the emission of a $W$ boson ($T\rightarrow bW$ and $B\rightarrow tW$). Likewise, the production of a single final-state vector-like $B$ quark occurs by virtue of the $tW\rightarrow B$ and $bZ\rightarrow B$ vertices, and similarly for the top quark's partner, $T$.
 
The kinematic properties of signal events are inferred from a phenomenological model of single VLQ production \cite{Matsedonskyi:2014mna, PhysRevD.101.115027, Buchkremer:2013bha} where both the left- and right-handed components of the VLQ mix with third-generation SM quarks via Yukawa couplings, giving rise to the interaction vertices mentioned above. In this picture, the coupling constants for interactions between the vector-like $B$ quark and the $W$, $Z$ and $H$ bosons, which regulate both the production cross-section and decay width, are given by:
\begin{equation}\label{eqn:couplings}
c_W = \kappa \sqrt{\frac{2\xi_W}{\rho_W}}, \ \ \ \ \ \ \
c_Z = \frac{m_Z}{m_W} \times \kappa \sqrt{\frac{2\xi_Z}{\rho_Z}}, \ \ \ \ \ \ \
c_H = \frac{1}{2} \frac{g_Wm_B}{m_W}\times \kappa \sqrt{\frac{2\xi_H}{\rho_H}}, \ \ \ \ \ \ \
\end{equation}
in which $\rho_{W\!,Z\!,H}$ are dimensionless kinematic factors approximately equal to 1 for $m_B > 1\text{ TeV}$ and $\xi_{W\!,Z\!,H}$ are dimensionless constants determining the coupling hierarchy and summing to unity. Furthermore, all three coupling constants scale with the universal coupling strength $\kappa$.
 
The various theoretically motivated multiplet scenarios are reflected in the choice of values for the $\xi$ constants: for a $B$ singlet, $\xi_W = 0.5$ and $\xi_Z = \xi_H = 0.25$; for a ($T,B$) doublet, $\xi_W = \xi_Z = 0.5$ and $\xi_H = 0$; and for a ($B,Y$) doublet, $\xi_W = 0$ and $\xi_Z = \xi_H = 0.5$.
In the asymptotic, high-$m_B$ limit, which holds to a good approximation down to $m_B \sim 1\text{ TeV}$, the values of $\xi$ in each multiplet state correspond to the branching fractions of the $B$ quark in the respective decay mode, and the resonance width can be expressed as a function of the coupling strength $\kappa$:
 
\begin{equation*}\label{eqn:kappa}
\Gamma_B \ \simeq \ \frac{g^2}{128\pi}\frac{m_B^3}{m_W^2}\times\kappa^2 \ \ \ \ \ \ \ \  \text{  for }m_B > 1\text{ TeV}
\end{equation*}

\begin{figure}[h]
\centering
\includegraphics[width=0.48\linewidth]{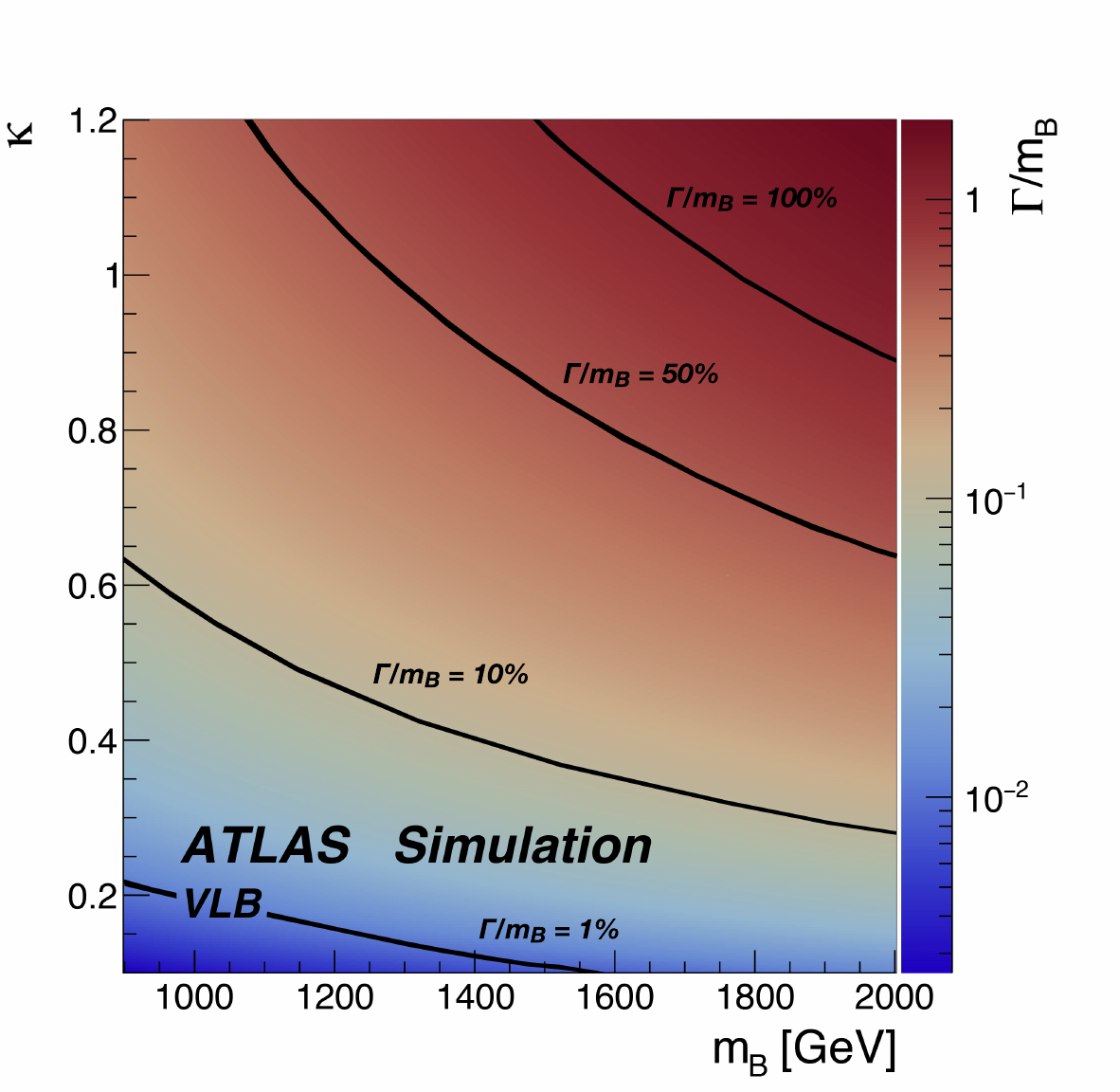}
\caption{Relative resonance width $\Gamma_B/m_B$  of a VLB as a function of the resonance mass and coupling strength $\kappa$. }\label{fig:Gm_mass_k}
\end{figure}

Figure~\ref{fig:Gm_mass_k} shows the values of the relative width of a vector-like  $B$ quark (VLB)  on a grid of values for the resonance mass and coupling strength.
 
\begin{figure}[h]{}
 
\centering
\subfloat[]{{\includegraphics[width=0.4\linewidth]{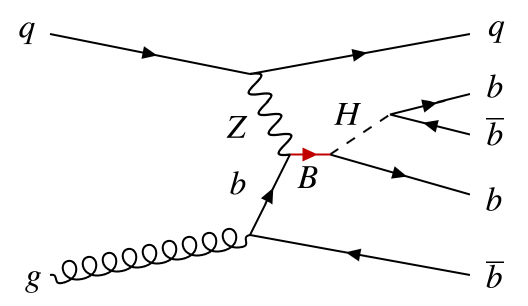}}\label{fig:s_zinit}}
\qquad
\subfloat[]{{\includegraphics[width=0.4\linewidth]{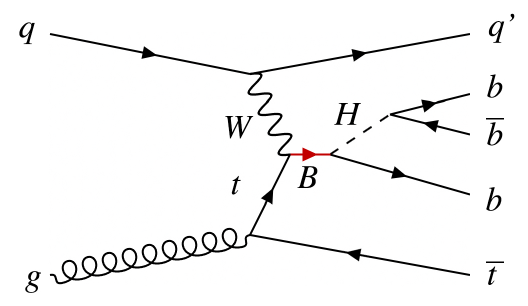}}\label{fig:s_winit}}
 
\subfloat[]{{\includegraphics[width=0.4\linewidth]{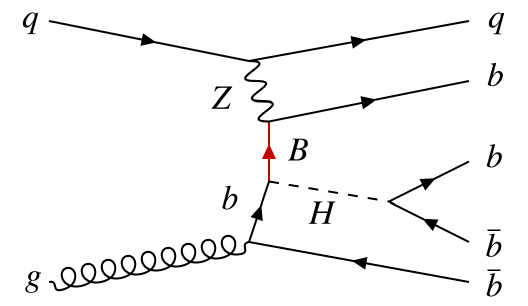}}\label{fig:t_zinit}}
\qquad
\subfloat[]{{\includegraphics[width=0.4\linewidth]{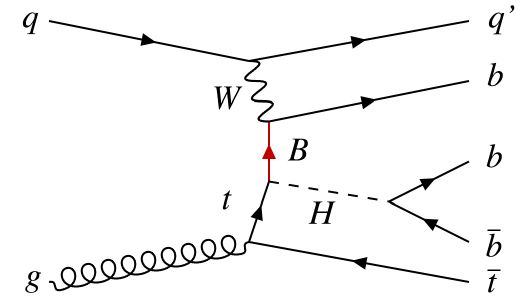}}\label{fig:t_winit}}\\
 
\caption{Feynman diagrams of the main leading-order $s$-channel (top row) and $t$-channel (bottom row) production modes of a single vector-like $B$ quark, as mediated by a $Z$ boson (a,c) or a $W$ boson (b,d).  The diagrams display the VLB decay resulting in the final state targeted by this search.}
\label{fig:feyn}
\end{figure}

This article details the search for a singly produced vector-like $B$ quark in the final state $B\to bH$ with $H\to b\bar{b}$, as shown in Figure~\ref{fig:feyn}.

A vector-like $B$ quark can be produced in the resonant $s$-channel (Figures~\ref{fig:s_zinit} and \ref{fig:s_winit}) as a result of either the electroweak interaction of an initial-state $b$-quark and a $Z$ boson, or of an initial-state $t$-quark and a $W$ boson, with the former process being the leading production mode for a vector-like $B$ singlet.
 
Conversely, strongly non-resonant single vector-like $B$ quark production arises through  $t$-channel processes as outlined by the diagrams in Figures~\ref{fig:t_zinit} and \ref{fig:t_winit}. This production mode, negligible when $\Gamma_B / m_B \le 10\%$, makes a sizeable contribution in large-width theoretical scenarios, but mostly results in low-mass off-shell $B$ quarks falling well outside the acceptance of the trigger selection employed by the analysis (see Section~\ref{sec:selection}).
 
In either production mode, the initial-state heavy-flavour quarks most often result from gluon splitting, and originate less frequently from the initial-state proton sea quarks because of the small values of the parton distribution functions (PDFs) for heavy flavour. The remaining heavy-flavour quark from the gluon splitting is generally outside of the momentum and rapidity acceptance of the ATLAS event reconstruction, and is not considered as a distinguishing feature of the signal events.
 
Several searches carried out on the \RunOne and early and full \RunTwo ATLAS datasets \cite{ATLAS:2022hnn,Aaboud:2018ifs,Aaboud:2018pii,EXOT-2016-35,ATLAS:2022ozf,ATLAS:2022tla} have targeted VLQ-compatible signatures. Searches for vector-like quark pair production using the full \RunTwo ATLAS dataset have excluded the presence of a vector-like $B$ quark occurring as a singlet (doublet) in the mass range $m_B < 1.2\ (1.3)$~\TeV. Searches by CMS targeting VLQ pair production with \RunTwo data resulted in compatible constraints on the theoretical model~\cite{CMS:2020ttz,CMS:2022fck}.
 
Production of vector-like $B$ quarks, both singly and in pairs, has been probed at CMS in the $bZ(b\bar{b})$ and $bH (b\bar{b})$ modes~\cite{Sirunyan:2017ezy,CMS-B2G-17-009}, and at ATLAS in the $bH(\gamma\gamma)$ final state, using $\sqrt{s} = 13$~\TeV\ data.
No evidence of singly produced vector-like $B$ quarks was found, and a 95\% confidence-level exclusion limit \cite{CLs_2002} was set on both the singlet hypothesis, for a branching fraction $\mathcal{B}(B\rightarrow bH$) $\approx 0.25$, and the $(B,Y)$ doublet hypothesis, for $\mathcal{B}(B\rightarrow bH$) $\approx 0.5$, ruling out masses below $\approx 1200$~\GeV.
 
This paper details the first search for a single vector-like $B$ quark in the $bH (b\bar{b})$ final state carried out on ATLAS data. The event selection is based on the presence of kinematic features most compatible with a vector-like $B$ signal as modelled by Monte Carlo simulations. Those features include the presence of a high-$p_{\mathrm{T}}$ large-radius ($R\ =\ 1.0$) jet opposite to a $b$-tagged high-$p_{\mathrm{T}}$ small-radius ($R\ =\ 0.4$) jet, the presence of two $b$-tagged track-jets matched to the large-$R$ jet, and the presence of one or more jets in the forward region of the detector.
Evidence of a signal is sought in the form of an excess in the reconstructed invariant-mass spectrum of the selected vector-like $B$ quark candidates, each formed by combining a large-$R$ jet and a small-$R$ jet satisfying the criteria outlined in Section~\ref{sec:reconstruction}.

A brief description of the ATLAS detector and an overview of the data and Monte Carlo samples employed in the analysis are provided in Sections~\ref{sec:detector} and~\ref{sec:samples} respectively. The reconstruction criteria for the physical objects involved in the analysis are outlined in Section~\ref{sec:reconstruction}, followed in Section~\ref{sec:selection} by an overview of the analysis-specific selection criteria used to maximise the search's sensitivity to the targeted signature. The next sections detail the procedure used to model the Standard Model background (Section~\ref{sec:background}), the treatment of the systematic uncertainties affecting the search (Section~\ref{sec:syst}) and the statistical framework (Section~\ref{sec:stats}) utilised  to interpret the search results (Section~\ref{sec:result}). Finally, the conclusions are given in Section~\ref{sec:conclusion}.


\section{ATLAS detector}
\label{sec:detector}

 
The ATLAS detector~\cite{PERF-2007-01} is a general-purpose particle detector used to investigate a broad range of physics processes. It includes inner tracking devices surrounded by a 2.3~m diameter superconducting solenoid, electromagnetic (EM) and hadronic calorimeters, and a muon spectrometer with a toroidal magnetic field. The inner detector consists of a high-granularity silicon pixel detector, including the insertable B-layer~\cite{ATLAS-TDR-2010-19,PIX-2018-001} installed after \RunOne of the LHC, a silicon strip detector, and a straw-tube tracker. It is immersed in a 2~T axial magnetic field and provides precision tracking of charged particles with pseudorapidity $|\eta|<2.5$.\footnote{ATLAS uses a right-handed coordinate system with its origin at the nominal interaction point (IP) in the centre of the detector. The positive $x$-axis is defined by the direction from the IP to the centre of the LHC ring, with the positive $y$-axis pointing upwards, while the beam direction defines the $z$-axis. Cylindrical coordinates $(r,\phi)$ are used in the transverse plane, $\phi$ being the azimuthal angle around the $z$-axis. The pseudorapidity $\eta$ is defined in terms of the polar angle $\theta$ by $\eta=-\ln\tan(\theta/2)$. The transverse momentum ($p_{\mathrm{T}}$) is defined relative to the beam axis and is calculated as $p_{\mathrm{T}} = p \sin(\theta)$.} The calorimeter system covers $|\eta|<4.9$. To measure EM showers, it contains finely segmented lead/liquid-argon (LAr) sampling calorimeters for $|\eta|<3.2$, and copper/LAr modules for higher $|\eta|$. A steel/scintillator hadronic calorimeter is used for $|\eta|<1.7$, complemented by copper/LAr endcaps and forward tungsten/LAr modules for $1.5<|\eta|<4.9$.  Outside the calorimeters, the muon system incorporates multiple layers of trigger and tracking chambers within a magnetic field produced by three superconducting toroids, enabling an independent precise measurement of muon track momenta for $|\eta|<2.7$. A dedicated trigger system is used to select events~\cite{TRIG-2016-01}. The first-level trigger is implemented in hardware and uses the calorimeter and muon detectors to accept events of interest at a rate below 100~kHz. This is followed by a software-based high-level trigger, which further reduces the rate to 1~kHz on average. An extensive software suite~\cite{ATL-SOFT-PUB-2021-001} is used in data simulation, in the reconstruction and analysis of real and simulated data, in detector operations, and in the trigger and data acquisition systems of the experiment.


\section{Data and simulated samples}
\label{sec:samples}

This analysis uses data from 139~fb$^{-1}$ of proton--proton (\textit{pp}) collisions at $\sqrt{s} = 13$~\TeV\ collected by ATLAS during the LHC's \RunTwo from 2015 to 2018. The data were collected during stable beam conditions with all relevant detector systems functional and producing good quality data \cite{Aad:2019hmz}. Events were selected online through a trigger signature that requires a single anti-$k_t$ jet~\cite{Cacciari:2008gp} with radius parameter $R = 1.0$ (large-$R$ jet) to satisfy transverse energy ($E_\mathrm{T}$) thresholds of 420~\GeV\ and 460~\GeV\ in the 2015--16 and 2017--18 data-taking periods, respectively. This trigger requirement is ${>}99$\% efficient for events passing the offline analysis selection of a large-$R$ jet with transverse momentum (\pt) over 480~\GeV.
 
The VLB signal is modelled by means of a Monte Carlo (MC) simulation based on the phenomenological Lagrangian outlined in Section~\ref{sec:intro}. Signal samples were generated with \MGNLO~\cite{Alwall:2011uj}, using the four-flavour scheme and the leading-order (LO) \NNPDF[2.3] PDF sets~\cite{Ball:2012cx}. Parton showering and hadronisation was handled by \PYTHIA[8.212]~\cite{Sjostrand:2014zea}, which used a set of tuned parameters called the A14 tune~\cite{ATL-PHYS-PUB-2014-021} for the underlying event. The \EVTGEN[1.2.0]~\cite{Lange:2001uf} program was used to model the properties of $b$-hadron decays. The detector response was simulated with \GEANT~\cite{Agostinelli:2002hh} and the events were processed with the same reconstruction software as that used for data~\cite{Aad:2010ah}. All simulated samples include the effects of multiple \textit{pp} interactions per bunch-crossing (\pileup), as well as the effect on the detector response due to interactions from bunch crossings before or after the one containing the hard interaction.
 
The samples were generated with the resonance pole mass ranging from $1~\TeV$ to $2~\TeV$, in steps of $200~\GeV$. To facilitate the statistical interpretation of the search results as inferences about the values of the coupling constants $c_W$, $c_Z$ and $c_H$, an event-by-event reweighting mechanism was introduced at the event-generation stage~\cite{Mattelaer:2016gcx} to simulate how different coupling strengths, ranging from $\kappa=0.1$ to $\kappa=1.6$ in Eq.~(\ref{eqn:couplings}), affect the properties of signal events.
The $Z$-initiated and $W$-initiated production modes giving rise to a single VLB quark (as shown in Figure~\ref{fig:feyn}) are treated separately, with independent samples generated for each mode. All samples were generated with a nominal coupling strength $\kappa = 0.4$, and inclusively with respect to Higgs boson decay mode. Interference effects between VLB production diagrams and SM diagrams resulting in the same final state were not taken into account in the simulation.
 
Signal samples are normalised to cross-section values calculated at leading order, assuming the four-flavour scheme and a singlet configuration. The normalisation is later corrected by means of next-to-leading-order (NLO) $K$-factors, assuming the narrow-width approximation and a five-flavour PDF scheme \cite{jhep:singProdTheo}.
 
The theoretical calculations of the single-$B$ production cross-section in the  $bH$ channel are considered unreliable in the very large width scenario. Consequently, the interpretation of the search results is only shown for configurations resulting in $\Gamma_B/m_B \le 50\%$ \cite{jhep:singProdTheo}.
 
Since the analysis targets a fully hadronic final state, multijet production is overwhelmingly the dominant source of background and a fully data-driven background estimation is performed, as described in Section~\ref{sec:background}. As a cross-check, simulated $t\bar{t}$ samples were also studied to confirm that the small (few percent) contribution from this process can be fully accounted for by the data-driven background estimation technique. The contribution from top quark pair production is modelled at NLO by the  \POWHEGBOX[v2]~\cite{Nason:2004rx,Frixione:2007vw,Alioli:2010xd} generator equipped with the \NNPDF[3.0nlo]~\cite{Ball:2014uwa}
PDF set for matrix-element calculations.
The top quark pair-production cross-section is scaled to a next-to-next-to-leading-order calculation in QCD
including resummation of
next-to-next-to-leading logarithmic soft gluon terms with \TOPpp[2.0]~\cite{Cacciari:2011hy,Beneke:2011mq,Baernreuther:2012ws,Czakon:2012zr,Czakon:2011xx}.
Parton showering, hadronisation and the underlying event were
simulated using \PYTHIA[8.230]~\cite{Sjostrand:2014zea} with the
\NNPDF[2.3lo]~\cite{Ball:2012cx} PDF set
and the A14 tune~\cite{ATL-PHYS-PUB-2014-021}.
 
The contributions to the total SM background from other processes, such as $Z$\,+\,jets production, are estimated to be negligible because of their small cross-sections and the low expected acceptance of the event selection for such processes.


\section{Object reconstruction}
\label{sec:reconstruction}

This search targets the production and decay of a VLB with a mass in the range of 1--2~\TeV, resulting in a final state composed of a high-$p_{\mathrm{T}}$ Higgs boson decaying into $b\bar{b}$, an energetic jet from the $b$-quark originating directly from the VLB decay, and an additional, softer, forward jet from the spectator quark involved in the hard scatter (as shown in Figure~\ref{fig:feyn}). The boosted Higgs boson decay system is reconstructed as a single large-radius (large-$R$) jet displaying a two-pronged energy profile, which originates from the hadronisation of the $b$- and $\bar{b}$-quarks. Conversely, the 	\textit{b}-quark from the initial VLB decay and the spectator quark, both expected features of a signal-like event, are reconstructed as standard, small-radius (small-$R$) jets. The identification of three $b$-hadrons in this topology is key to suppressing the background.
 
Events are checked in order to remove those with noise bursts or coherent noise in the calorimeters, as well as those containing large energy deposits from non-collision or cosmic sources of background. Collision vertices are reconstructed from inner-detector tracks with $p_{\mathrm{T}}>0.5$~\GeV. The primary vertex in each event is chosen to be the one with the largest sum of the squared transverse momenta of all associated tracks. Events without a reconstructed primary vertex are rejected.

Events containing isolated, charged leptons (electrons or muons) are removed in this analysis, since no leptons are expected in the final state under study. This applies to events containing electron candidates  with $E_{\mathrm{T}}>25$~\GeV\ that satisfy the \enquote{loose} identification criteria defined in Ref.~\cite{EGAM-2018-01} or muons with $p_{\mathrm{T}}>25$~\GeV\ satisfying the \enquote{medium} quality requirements~\cite{ATLAS:2020auj}. Also, to ensure orthogonality to VLB searches targeting the $H\to\gamma\gamma$ channel, events with isolated photons that meet the \enquote{tight} identification criteria~\cite{EGAM-2018-01} are removed if a pair of photons has an invariant mass in the range $105{-}160$~\GeV.
 
Small-$R$ jets are reconstructed by applying the \antikt algorithm \cite{Cacciari:2008gp}, with radius parameter $R=0.4$, to inner-detector tracks associated with the primary vertex and calorimeter energy clusters selected through a particle-flow reconstruction algorithm~\cite{PERF-2015-09}. An appropriate  energy  calibration is applied to both the input clusters \cite{PERF-2011-03} and the final reconstructed jet \cite{nowPublic}. Additionally, a \pileup subtraction procedure \cite{PERF-2016-04} is applied along with a global sequential calibration to account for flavour dependencies. To suppress jets arising from \pileup, a jet-vertex-tagging (JVT) technique using a multivariate likelihood~\cite{PERF-2014-03} is applied to jets with $\pt<60$~\GeV, ensuring that selected jets are matched to the primary vertex via their associated tracks. Jets with $|\eta|>2.4$, falling outside the inner-detector acceptance, undergo a tighter selection via a specially designed and trained forward-JVT algorithm~\cite{PERF-2016-06}. 
 
Large-$R$ jets are built by applying the anti-$k_t$ algorithm with radius parameter $R=1.0$ to three-dimensional topological clusters of energy deposits in the calorimeter that are calibrated to the hadronic energy scale with the local cluster weighting (LCW) procedure~\cite{PERF-2014-07}.
The reconstructed jets are \enquote{trimmed}~\cite{Krohn:2009th} to reduce contributions from \pileup and soft interactions. This is done by reclustering the jet constituents into subjets, using the $k_t$ algorithm~\cite{Ellis:1993tq,CATANI1993187} with a radius parameter $R=0.2$, and discarding subjets with \pt less than 5\% of the parent jet \pT~\cite{ATLAS-CONF-2015-035}. The large-$R$ jet four-momentum is then recomputed from the four-momenta of the remaining subjets and corrected using simulation~\cite{PERF-2011-03,PERF-2012-01}.
 
Small-$R$ jets in the range $|\eta_j|<2.5$ that contain a $b$-hadron are recognised (\enquote{$b$-tagged}) using the \enquote{DL1r} algorithm~\cite{FTAG-2018-01}. This algorithm is based on a multivariate classification technique that uses an artificial deep neural network to combine information about the impact parameters of tracks and the topological properties of secondary and tertiary decay vertices reconstructed from tracks associated with the jet. The analysis selects $b$-jets by using the DL1r working point which has an efficiency of 70\% for identifying true $b$-jets in simulated SM \ttbar events. The corresponding mis-tagging efficiency for $c$-jets (containing $c$-hadrons) and light-flavour jets is estimated to be 10\% and 0.2\%, respectively.
 
In order to explore each large-$R$ jet for the presence of one or more $b$-hadrons, which would be expected in boosted $H\rightarrow b\bar{b}$ decays, variable-radius (VR) track-jets are matched to the large-$R$ jet via \enquote{ghost association} \cite{PERF-2012-02, Cacciari:2007fd, Cacciari:2008gn} and subsequently inspected for $b$-tagging. The track-jets are built from inner-detector tracks by using the anti-$k_t$ algorithm with a radius parameter $R$ inversely proportional to the jet \pt~\cite{Krohn:2009zg}:
\begin{equation}\nonumber
R \to R_{\text{eff}}(p_{\mathrm{T}})=\rho/p_{\mathrm{T}}\,.
\end{equation}
The $\rho$-parameter, which controls the effective radius $R_{\text{eff}}$, is set to $\rho=30$~\GeV. Two additional parameters, $R_{\text{min}}$ and $R_{\text{max}}$, are used to place lower and upper bounds on $R_{\text{eff}}$, and these are set to 0.02 and 0.4, respectively~\cite{ATL-PHYS-PUB-2017-010}. The values of these parameters were chosen by examining the efficiency of identifying two $b$-jets within a large-$R$ jet associated with a high-\pt Higgs boson decaying into a $b$-quark pair~\cite{PERF-2017-04}.
Similarly to the tagging of small-$R$ jets, a version of the \enquote{DL1} algorithm was specifically retrained to $b$-tag VR track-jets. For a 70\% efficiency to identify true $b$-jets as measured in simulated SM \ttbar events, the mis-tag efficiency for $c$-jets and light-flavour jets is about 10\% and 0.25\%, respectively.
For both the small-$R$ jets and track-jets, the efficiencies of identifying $b$-jets, $c$-jets, and light-flavour jets are corrected in the simulation to account for deviations from the efficiencies observed in data~\cite{FTAG-2018-01}.
 
Events are vetoed if any of the track-jets inspected for $b$-tagging is found to be collinear with any other track-jet with $\pT > 10$~\GeV\ in the event. For this cleaning procedure, track-jet collinearity is defined by the angular separation $\Delta R = \sqrt{(\Delta \eta)^2+(\Delta \phi)^2}$ between the two examined track-jets being smaller than both of the effective jet radii. The collinearity veto prevents events with ambiguous track to track-jet matching, and therefore uncalibrated flavour-tagging performance, from entering the analysis, at the cost of an observed 6\% signal efficiency loss across the accessible resonance mass spectrum.


\section{Event selection and categorisation}
\label{sec:selection}

As mentioned above, Higgs boson candidates (HC) are reconstructed as single large-$R$ jets. Events are selected if they have at least one eligible HC reconstructed as a large-$R$ jet with $\pt^\text{HC} > 480$~\GeV, $|\eta|<2.0$ and at least two matched track-jets with $\pt>50$~GeV.  The transverse momentum requirement on the large-$R$ jet matches the beginning of the full efficiency plateau of the lowest-threshold unprescaled large-$R$ jet trigger selected to define the analysis dataset.
 
Reconstructed Higgs boson candidates are classified according to their $b$-tagged track-jet multiplicity, allowing candidates with a higher number of $b$-tagged jets to be prioritised if multiple eligible HCs are present within a single event. The two HC categories are labelled \enquote{\texttt{H2T2B}}, for candidates with two matched $b$-tagged track-jets, and \enquote{\texttt{H2T1B}} otherwise. The presence of at least one $b$-tagged track-jet matched to the large-$R$ jet is required for Higgs boson candidate eligibility. In the rare cases (less than 1\%) where more than one eligible HC is reconstructed, the candidate with the highest mass is selected.
 
\begin{figure}[h]
\centering
\subfloat[ ]{{\includegraphics[width=0.4\linewidth]{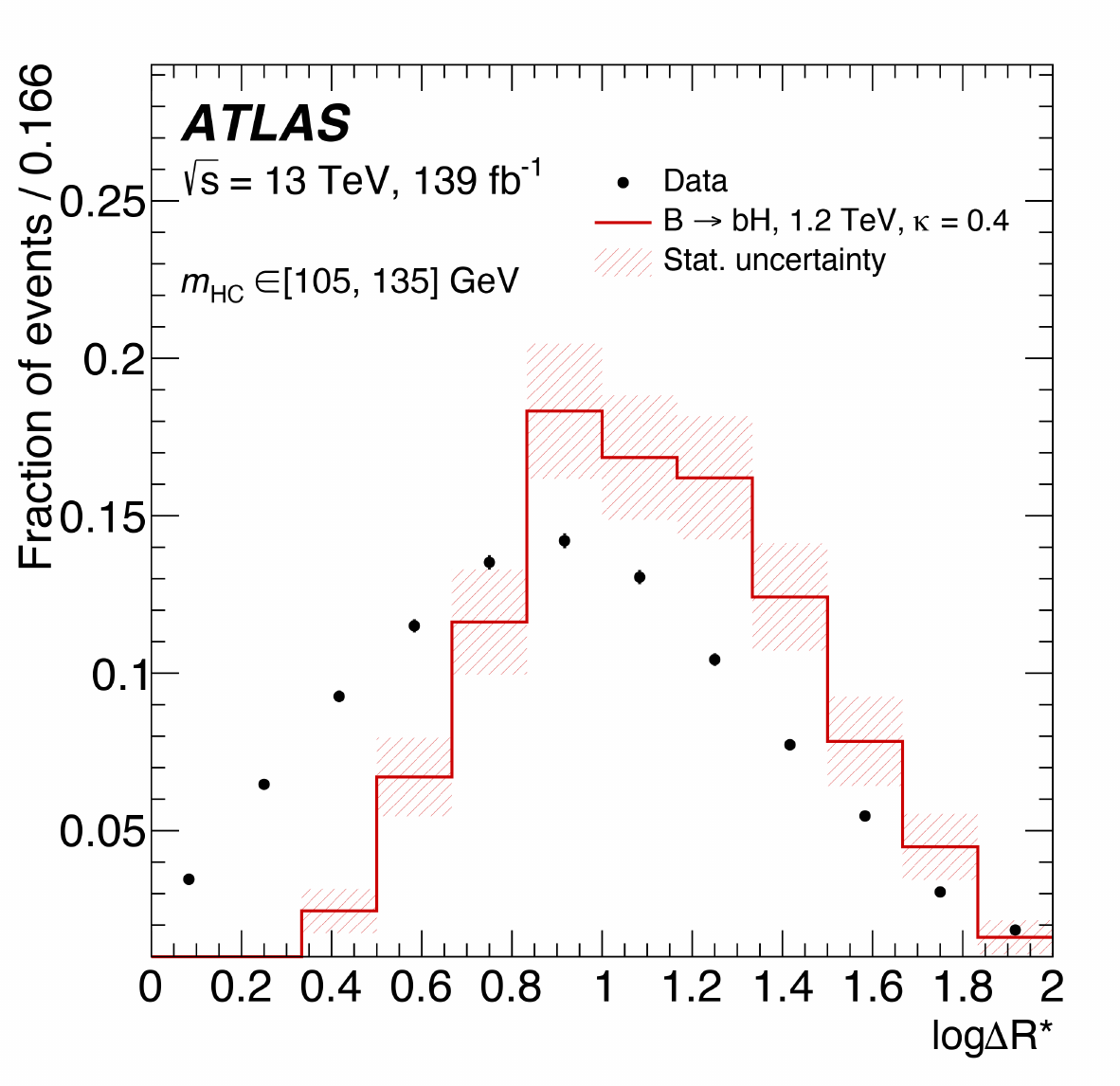}}\label{fig:sb_dRLogNorm_peak_1200}}
\qquad
\subfloat[ ]{{\includegraphics[width=0.4\linewidth]{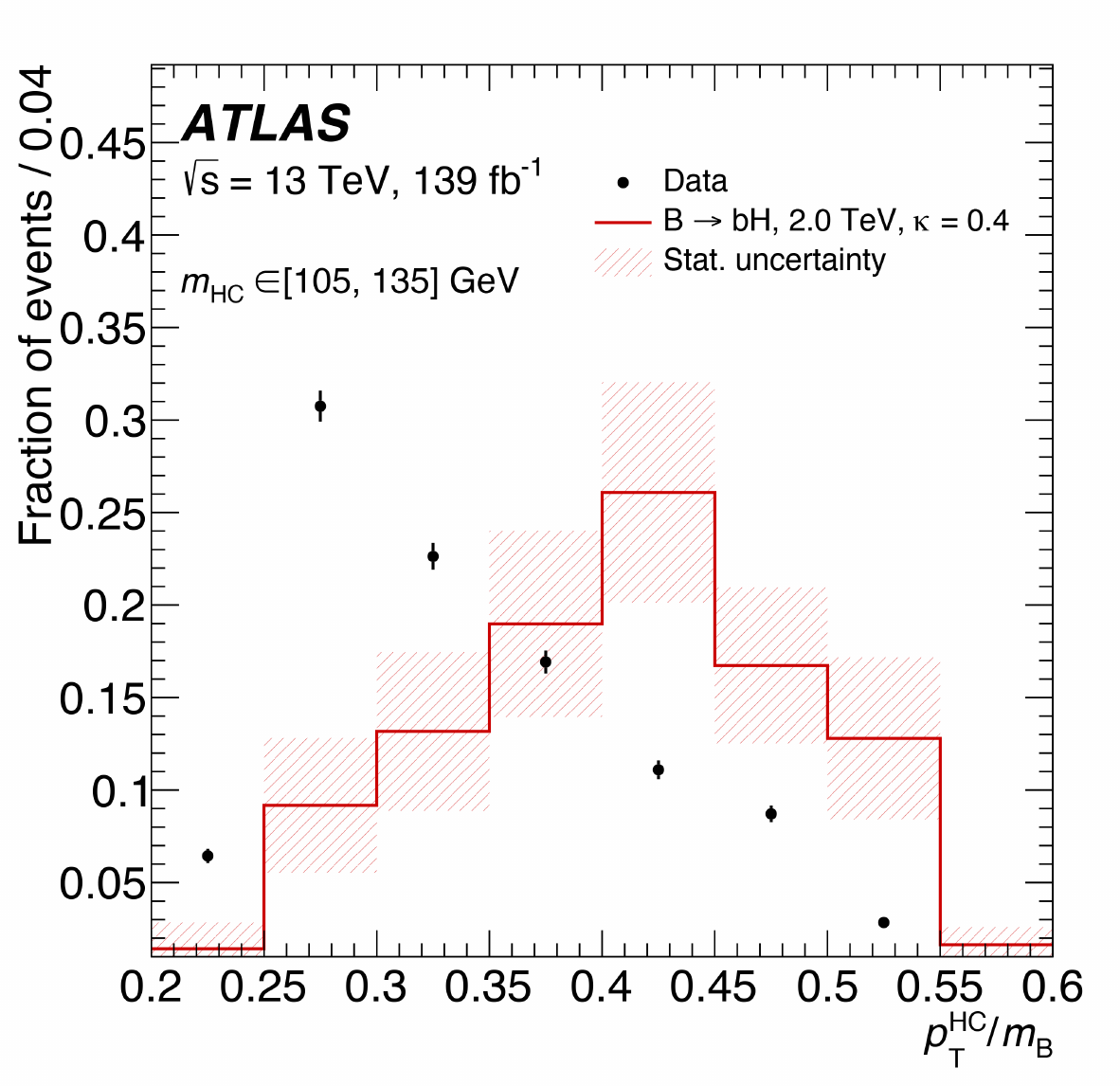}}\label{fig:sb_hOverm_peak_2000}}
 
\caption{(a) $\log{\!\Delta R^*}$ distributions for the data and simulated 1.2~\TeV\ VLB signal; and (b)  $\pt^\text{HC}/m_{B}$ distributions for the data and simulated 2~\TeV\ VLB signal. The displayed events belong to a 150-\GeV-wide window centred on the signal resonance pole mass. All histograms are normalised to the same area for an easier comparison of the shapes. The error bars on the data points refer to the statistical uncertainties only, while the shaded error bands on the signal distribution refer to the statistical uncertainty of the MC simulation.}
\label{fig:sb_dRLogNorm_peak}
\end{figure}

VLB candidates are formed by combining a HC with a $b$-tagged small-$R$ jet required to have $\pt > 400$~\GeV, $|\eta|<2.5$ and an angular distance $\Delta R > 2.0$ from the HC.
Two further selection criteria are applied to exploit the subjet structure of HCs in signal events and the correlation between $\pt^\text{HC}$ and the VLB candidate mass, $m_{B}$. The first is captured by the quantity $\log{\!\Delta R^*}$, defined as:
\begin{equation}\nonumber
\log{\!\Delta R^*} \ = \
\log{\Bigg[\frac{\Delta R\Big(\text{tj0, tj1}\Big)}{\min{\Big[R_\text{eff}^\text{tj0}, R_\text{eff}^\text{tj1}\Big]}}\Bigg]},
\end{equation}
where  tj0 and tj1 are the two highest-\pt\ track-jets associated with the HC, and $R_\text{eff}^\text{tj0}$ and $R_\text{eff}^\text{tj1}$ are their \pt-dependent effective radii.
The second quantity is the ratio of the HC \pt to the reconstructed VLB invariant mass, $\pt^\text{HC}/m_{B}$, which is used to reject events where the two leading jets are produced with a large value of $\Delta \eta$, a configuration known to be prevalent in high-\pt multijet production. The distributions of $\log{\!\Delta R^*}$ and $\pt^\text{HC}/m_{B}$ in signal and background events are shown in Figure~\ref{fig:sb_dRLogNorm_peak}. The behaviour of the variables under study in the SM background is approximated by a fully orthogonal, signal-depleted data sample where the small-$R$ jet is required to not be $b$-tagged.
Events are selected if they satisfy $\log{\!\Delta R^*}>0.67$ and $\pt^\text{HC}/m_{B}>0.4$. If more than one VLB candidate in an event fulfils the event selection requirements at this point, which occurs in approximately in 2\% of the events where a VLB candidate is found, the candidate with the lowest $\pt^\text{B}/m_{B}$ is chosen because the search targets a low-$\pt$, high-mass decaying particle.
Finally, since the signal event topology involves a forward spectator quark, events are required to have at least one jet with $\pt>40$~\GeV\ and $|\eta|>2.5$.
 
The search is restricted to a data subsample of maximal signal purity by using a signal region (SR) defined by the mass of the HC, required to be between 105~\GeV\ and 135~\GeV, and its $b$-tagging category, required to be \texttt{H2T2B}. The lower-purity sample of \texttt{H2T1B} events passing the full event selection and the HC mass requirement is preserved for the purpose of validating the background modelling procedure. A comprehensive breakdown of all preselection, event reconstruction and kinematic selection criteria is provided in Table~\ref{tab:sel}.
 
The full event selection efficiency for simulated signal events varies as a function of the coupling strength $\kappa$ regulating the resonance width. For a benchmark signal model with $m_B = 1200$~\GeV\ and $\kappa = 0.4$, approximately 4\% of all simulated $B\rightarrow bH$ events (inclusive with respect to the Higgs decay) have one eligible VLB candidate, but only 0.7\% eventually pass the kinematic selection outlined above and enter the signal region. The main factors affecting the reconstruction and selection efficiency are the large-$R$ jet $\pt$ threshold, set at 480~\GeV\ to ensure 100\% trigger efficiency, the triple-$b$-tagging efficiency and the requirement of signal region events to feature at least one jet in the forward region of the detector. The overall selection efficiency rises slightly with increasing $m_B$, as more events on average satisfy the transverse momentum requirement on the leading jet.
 
\begin{table}
\centering
\caption{Summary of all preselection, reconstruction and kinematic selection steps leading to the full definition of the signal region and a number of orthogonal control data samples that are used for validation purposes.}
\label{tab:sel}
\begin{tabular}{|C{11mm}|C{11mm}|C{11mm}|C{11mm}|C{11mm}|C{11mm}|C{11mm}|C{11mm}|}
\hline
\hline
\hline
\multicolumn{8}{|c|}{\textbf{Preselection}}\Tstrut\\[0.5ex]
\hline
\hline
\hline
\multicolumn{8}{|c|}{$\geqslant 1$ large-$R$ jet, $\pt > 480$~\GeV}\Tstrut\\[0.5ex]
\hline
\multicolumn{8}{|c|}{No leptons \& no $\gamma\gamma$ pairs with $m_{\gamma\gamma}\in [105, 160]$~\GeV}\Tstrut\\[0.5ex]
\hline
\multicolumn{8}{|c|}{$\geqslant 2$ track-jets associated with the large-$R$ jet, \ $\geqslant$ 1 $b$-tagged track-jet}\Tstrut\\[0.5ex]
\hline
\multicolumn{8}{|c|}{$\geqslant 1$ small-$R$ jet with $\pt > 300$~\GeV}\Tstrut\\[0.5ex]
\hline
\multicolumn{8}{|c|}{$\Delta R(\text{small-}R\text{\ jet}, \text{large-}R\text{\ jet)} > 2.0$}\Tstrut\\[0.5ex]
\hline
\hline
\hline
\multicolumn{8}{|c|}{\textbf{HC reconstruction}}\Tstrut\\[0.5ex]
\hline
\hline
\hline
\multicolumn{8}{|c|}{Any large-$R$ jet with $\pt > 480$~\GeV}\Tstrut\\[0.5ex]
\hline
\multicolumn{8}{|c|}{$\geqslant 2$ ghost-matched track-jets with $\pt > 50$~\GeV}\Tstrut\\[0.5ex]
\hline
\multicolumn{8}{|c|}{Pass collinearity veto}\Tstrut\\[0.5ex]
\hline
\multicolumn{4}{|c|}{Highest $b$-tag multiplicity: 2 track-jets} &
\multicolumn{4}{c|}{Highest $b$-tag multiplicity: 1 track-jet}\Tstrut\\[0.5ex]
\hline
\multicolumn{8}{|c|}{Select candidate with largest $m_{\text{HC}}$}\Tstrut\\[0.5ex]
\hline
\hline
\hline
\multicolumn{8}{|c|}{\textbf{VLB candidate reconstruction}}\Tstrut\\[0.5ex]
\hline
\hline
\hline
\multicolumn{8}{|c|}{HC + small-$R$ jet, $\pt$(small-$R$ jet) > 400~\GeV}\Tstrut\\[0.5ex]
\hline
\multicolumn{8}{|c|}{$\Delta R(\text{small-}R\text{\ jet}, \text{large-}R\text{\ jet})$ > 2.5}\Tstrut\\[0.5ex]
\hline
\hline
\hline
\multicolumn{8}{|c|}{\textbf{Kinematic selection}}\Tstrut\\[0.5ex]
\hline
\hline
\hline
\multicolumn{8}{|c|}{$\log\Delta R^{*}$ > 0.67}\Tstrut\\[0.5ex]
\hline
\multicolumn{8}{|c|}{$\pt^\text{HC}/m_{B}$ > 0.4}\Tstrut\\[0.5ex]
\hline
\multicolumn{8}{|c|}{$m_{\text{HC}} \in [105, 135]\text{ GeV}$}\Tstrut\\[0.5ex]
\hline
\multicolumn{2}{|c|}{$\geqslant 1$ forward jet}\Tstrut & \multicolumn{2}{c|}{$=\ 0$ forward jet} & \multicolumn{2}{c|}{$\geqslant 1$ forward jet}\Tstrut & \multicolumn{2}{c|}{$=\ 0$ forward jet} \Tstrut\\[0.5ex]
\hline
\hline
\hline
\multicolumn{8}{|c|}{\textbf{Small-$R$ jet $b$-tagging status}}\Tstrut\\[0.5ex]
\hline
\hline
\hline
Tag & No Tag & Tag & No Tag & Tag & No Tag & Tag & No Tag\Tstrut\\[0.5ex]
\hline
\hline
SR & \multicolumn{7}{c|}{Control samples} \Tstrut \\ [0.5ex]
\hline
\end{tabular}
\end{table}


\section{Data-driven background modelling}
\label{sec:background}

As mentioned in Section~\ref{sec:samples}, the background in this analysis is largely dominated (over 90\%) by \enquote{multijet} events featuring QCD  production of multiple jets in the final state, with most of the remainder traceable to $t$-quark pair production. The contribution to the SM background from  $Z$ boson production in association with jets is estimated from published results to be of the order of $1\%$~\cite{ATLAS:2020juj}.
 
\begin{figure}
\centering
\includegraphics[width=0.5\textwidth]{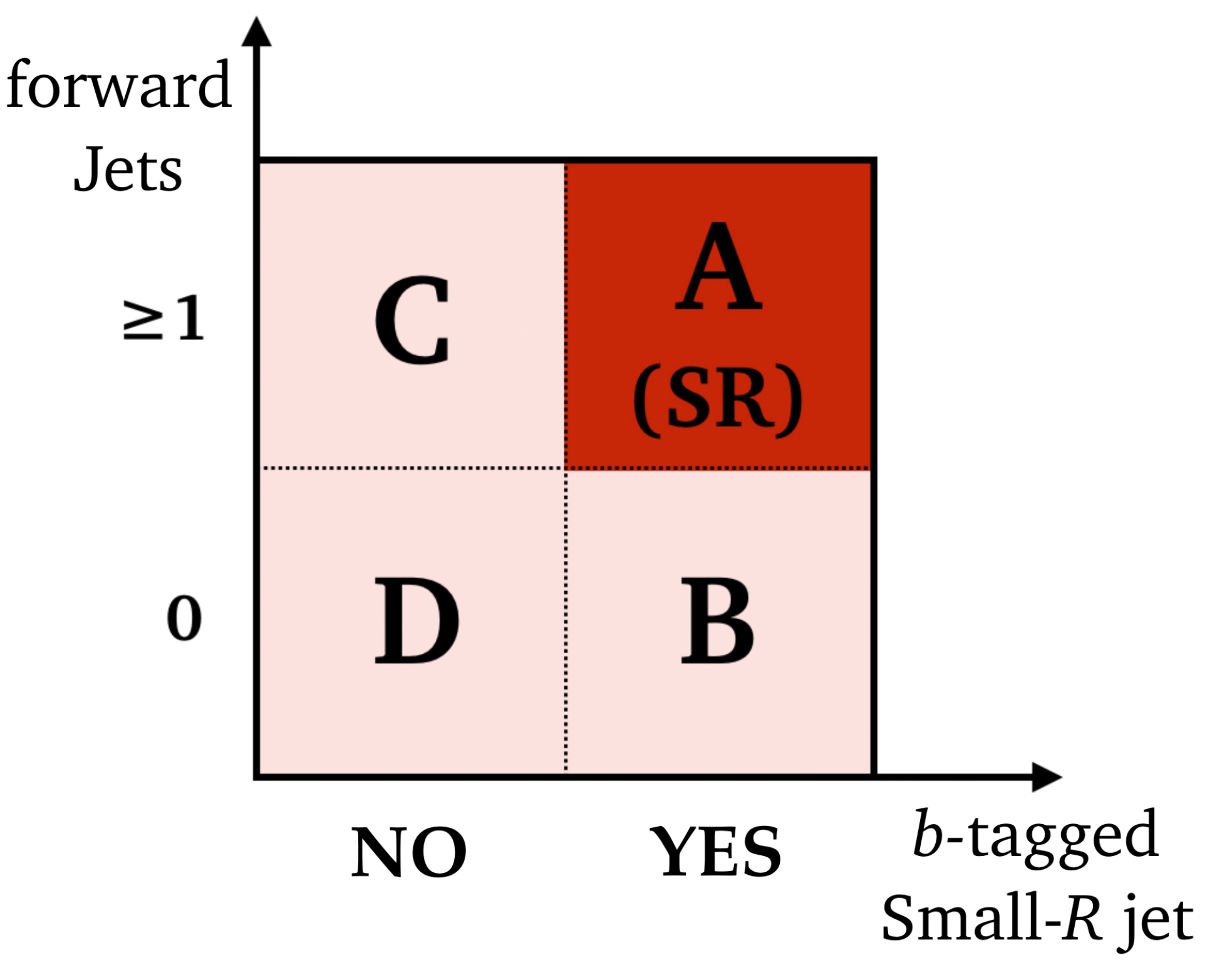}
\caption{Schematic representation of the ABCD partitioning employed in this search.}
\label{fig:abcd_new}
\end{figure}

Since it is challenging to include and properly model all the processes that could contribute to the multijet background in an MC simulation, a fully data-driven background estimation is performed, based on the well-established and often-used \enquote{ABCD} method. The shape of the $m_B$ distribution, which is used as the discriminating variable in the statistical analysis of the data, is likewise estimated through a modified ABCD-like procedure. The two event properties used to define the ABCD partitioning are (a) the $b$-tagging classification of the small-$R$ jet in the VLB candidate; and (b) the presence or absence of forward (abbreviated as \enquote{fwd}) jets in the event (see Figure~\ref{fig:abcd_new}). In this analysis, events passing the full selection described in Section~\ref{sec:selection} fall in region~A, while events lacking either a forward jet or a $b$-tagged small-$R$ jet in the VLB candidate, or both, fall in regions B, C, or D respectively.
For this ABCD partitioning, the A-region data sample, where $m_\text{HC}\in [105, 135]~\GeV$, is predicted to have maximal signal purity. Because of this, the A-region data events were temporarily removed, thereby \enquote{blinding} the analysis to the possible presence of a signal which could otherwise bias the SM background modelling.
 
Provided that the two properties used to separate the events into the ABCD regions are uncorrelated, the ratio of background events in regions A and B would be equal to that in regions C and D. From this identity, it follows that $\tilde{N}_\text{A}$, the expected number of background events in region A, can be estimated to be
$
\tilde{N}_\text{A} = k_\text{fwd}N_\text{B}
$,
where $k_\text{fwd}=N_\text{C}/N_\text{D}$.
 
While the two event variables used to define the ABCD partitioning are required to be as uncorrelated as possible, a level of residual correlation is expected, and accounted for by computing the value of the correlation-sensitive $\rcorr$ estimator, defined as:
\begin{equation}\nonumber
\rcorr= \frac{N_\text{A}}{N_\text{B}} \cdot \frac{N_\text{D}}{{N_\text{C}}}\,.
\end{equation}
 
The value of $\rcorr$ can only be computed in control samples where the data is not blinded in any ABCD region. Three such samples are defined in the present analysis. The first sample contains events selected with the same selection criteria as in Section~\ref{sec:selection}, except that the HC invariant mass is required to be in the range $m_\text{HC} \in [135, 200]$~\GeV, which defines the validation sideband (VS) mass window. This event sample is labelled \enquote{\texttt{H2T2B\_VS}}, to distinguish it from the main analysis sample, labelled \enquote{\texttt{H2T2B}}, where the HC falls in the Higgs mass window (signal region).
The other two control samples comprise events that satisfy the same selection criteria as \texttt{H2T2B} events, except that only one of the track-jets associated with the HC is $b$-tagged. These events are labelled \enquote{\texttt{H2T1B}}, with the \texttt{VS} label when falling in the Higgs mass sideband.
All these control samples have significantly lower sensitivity than the \texttt{H2T2B} sample, owing to their smaller signal content and larger expected background, and are estimated only to be sensitive to VLB production cross-sections that are already excluded by previous searches.
 
The values of $\rcorr$ computed in the three aforementioned control samples are consistent within their statistical uncertainties:
\begin{equation}\nonumber
\rcorr^\text{H2T2B\_VS} = 1.11\pm 0.05
\hspace{1cm}
\rcorr^\text{H2T1B\_VS}= 1.12\pm 0.02
\hspace{1cm}
\rcorr^\text{H2T1B}= 1.11\pm 0.03 \,.
\end{equation}
These values, consistently greater than one, imply a slight underestimation of the background yield by the uncorrected ABCD method as a result of the residual correlation between the ABCD axes. Consequently, a correction to the background prediction in region A of \texttt{H2T2B} is implemented by scaling the transfer factor $k_\text{fwd}$ by $\rcorr$ itself:
\begin{equation}\nonumber
\tilde{N}_\text{A} = \rcorr^\text{H2T2B\_VS} \cdot k_\text{fwd} \cdot N_\text{B}\,.
\end{equation}
 
\begin{figure}[htb]
\centering
\subfloat[]{{\includegraphics[width=0.4\linewidth]{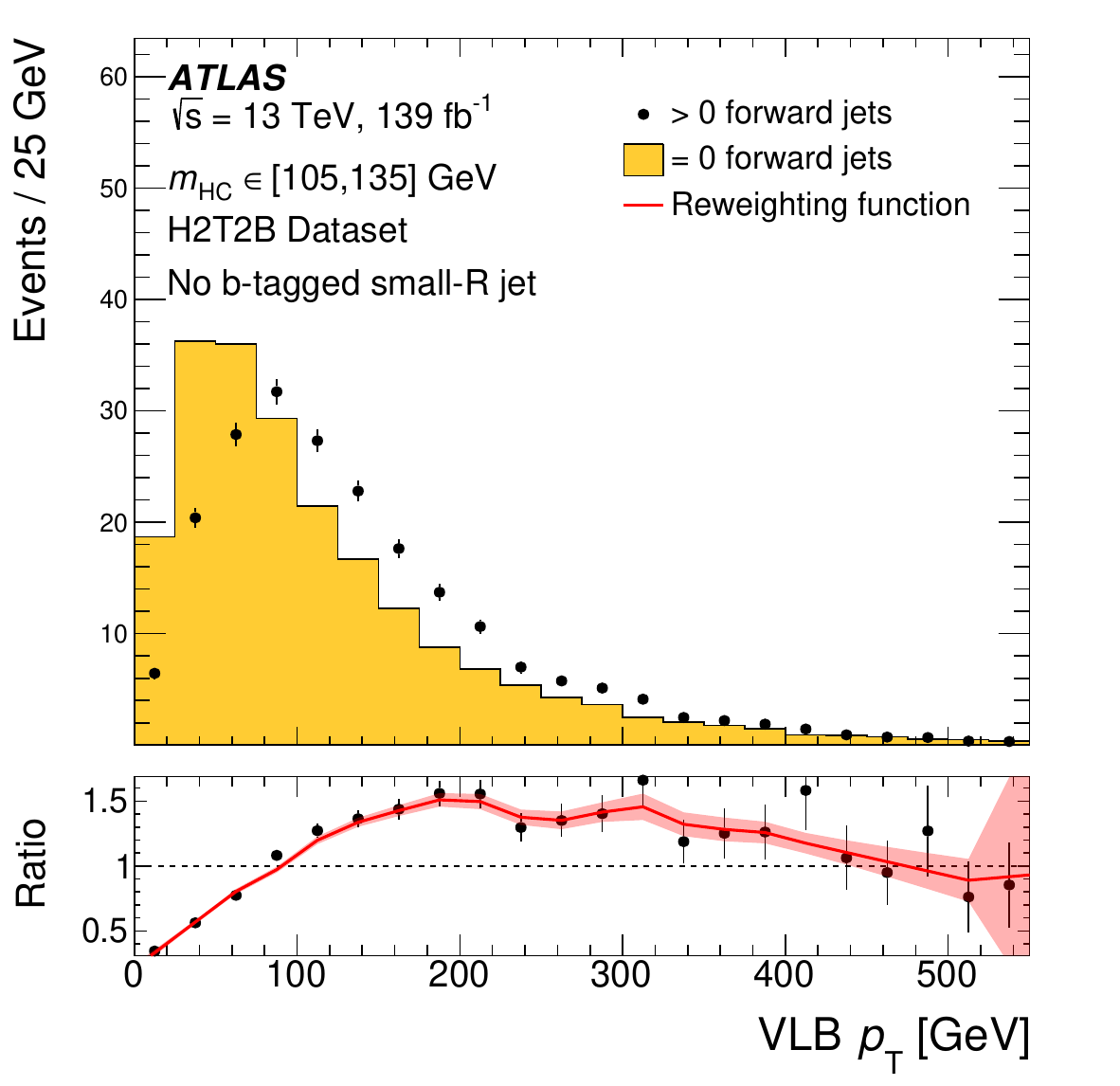}}\label{fig:ABCDWeight_vlqCand_pt}}
\qquad
\subfloat[]{{\includegraphics[width=0.4\linewidth]{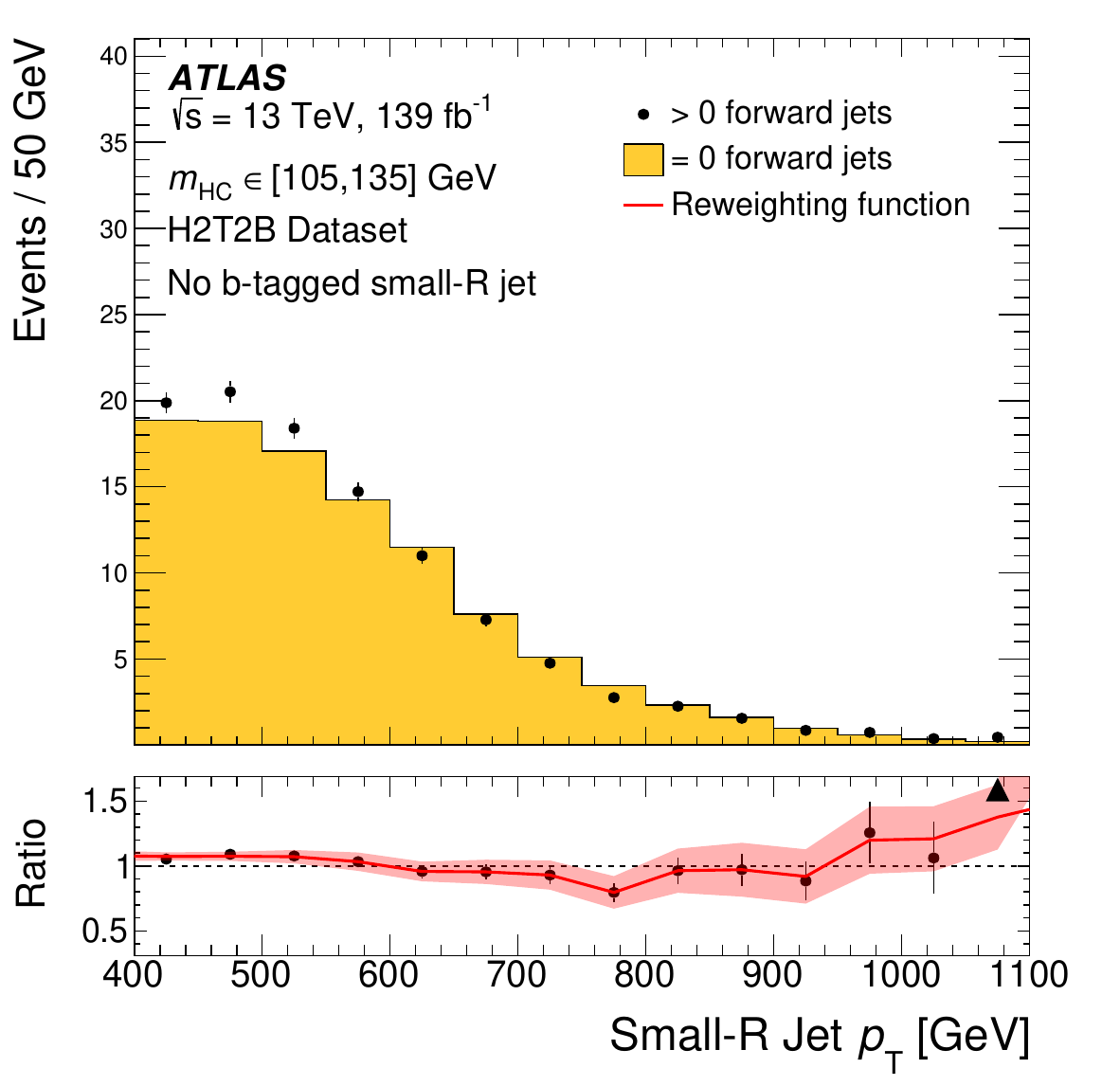}}\label{fig:ABCDWeight_vlqJet_pt}}\\
 
\caption{Comparison of the distributions in ABCD regions C and D for the two variables used to calculate the correction weights: (a) the VLB \pT  and (b) the small-$R$ jet \pT. All events shown belong to the Higgs mass window $m_\text{HC} \in [105, 135]~\GeV$, but only C-region events have one or more forward jets. The normalisation scale factor $k_\text{fwd}$ is already applied to D-region events. The non-parametric corrections and uncertainties, obtained by smoothing the ratio via Gaussian kernel regression, are shown with the red line and pink shaded area, respectively, in each of the ratio plots. The product of corrections in (a) and (b) are applied to B-region events.}
\label{fig:CD_weights}
\end{figure}

Additionally, in order to correctly predict the kinematic shapes in region A, the event sample in region B is reweighted, using per-event weights derived by comparing the distributions of two kinematic variables in regions C and D, to compensate for differences between events with no forward jets and events with one or more forward jets. The two variables chosen for this purpose are the VLB \pt\ and the \pt\ of the small-$R$ jet participating in the VLB reconstruction, which display the largest discrepancies between data distributions in regions C and D, as shown in Figure~\ref{fig:CD_weights}. The weights in each of the two variables are calculated from the bin-by-bin ratios of the normalised distributions, with non-parametric smoothing applied using Gaussian kernel regression~\cite{Nada:1964} to smooth out the effects of any statistical fluctuations. By construction, the reweighting procedure does not affect the yield of the region A background prediction. The template ratio regression method used to extract smooth, continuous event-weight functions comes naturally with an associated $\pm 1\sigma$ uncertainty band, as displayed in Figure~\ref{fig:CD_weights}.
The overall weights are the product of the weights extracted from the two kinematic variables. After this reweighting procedure all kinematic distributions in region D are found to be in good agreement with those in region C, indicating a satisfactory level of closure for the kinematic reweighting method.
The uncertainties in the event weights applied to each B-region event are propagated to the final background model for signal region data and taken as systematic uncertainties in the shape and yield of the predicted SM background in the signal region.
 
\subsection{Effect of signal contamination on the background model}\label{sub:cont_intro}
In an ideal scenario, the ABCD partitioning is tuned in such a way as to contain a large majority of signal events within the blinded target region, with only minimal spillage into the remaining three regions. A perfectly clean signal separation is, however, impossible in practice because of inefficiencies in flavour tagging and in the identification of the spectator quark as a forward jet.
This causes a non-negligible fraction of the predicted signal yield to fall into regions B, C and D, potentially affecting the modelling of the multijet background. A breakdown of the distribution of data and signal events across the ABCD plane is displayed in Table~\ref{tab:app_yield}.
\begin{table}[h]
\centering
\caption{Event yields in the four ABCD regions in the Higgs mass window ($m_\text{HC}\in[105,\,135]$~\GeV) in the reference $m_B = 1.3$~\TeV\ and $\kappa = 0.4$ isospin-singlet signal hypothesis (left) and in data (right). The expected data yield in the signal region (ABCD-region A), corresponding to the predicted SM background contribution, is displayed in boldface with its associated uncertainty.}
\label{tab:app_yield}
\begin{tabular}{c|c|c}
\multicolumn{3}{c}{Singlet, $m_B$ = 1.3 TeV, $\kappa$ = 0.4 } \Tstrut\\[0.6ex]
\diagbox{fwd jets}{$b$-tags} & 2 & 3  \TBstrut\\[0.5ex]
\hline
\hline
$\ge$ 1   & 19 & 26   \TBstrut\\[0.5ex]
\hline
$=$ 0  & 11 & 15 \TBstrut\\[0.5ex]
 
\end{tabular}
\qquad
\begin{tabular}{c|c|c}
\multicolumn{3}{c}{Data} \Tstrut\\[0.6ex]
 
\diagbox{fwd jets}{$b$-tags} & 2 & 3  \TBstrut\\[0.5ex]
\hline
\hline
$\ge$ 1   & \phantom{0}5310  & \textbf{257\,$\pm$\,25}  \TBstrut\\[0.5ex] 
\hline
$=$ 0 & 23800 & 972 \TBstrut\\[0.5ex] 
 
\end{tabular}
 
\end{table}
The signal spillage in regions C and D, well below 1\%, would have a negligibly small impact on $k_\text{fwd}$ (well below its associated statistical uncertainty). The number of signal events in region B is, however, non-negligible, and would translate into an overestimation of the background below a possible signal by approximately 15\% of the predicted A-region signal yield.\footnote{The background overestimation caused by signal spillage in region B is estimated as $N_\text{cont} \simeq N_\text{B}^\text{sig}\times k_\text{fwd}\times \rcorr$.} An MC-derived correction of the background model is implemented to correct the background estimate for this potential effect, as described in greater detail in Section~\ref{sec:stats}.
 
\subsection{Background model validation}\label{sub:validation}

The full background modelling procedure is applied to the previously defined \texttt{H2T2B\_VS} and \texttt{H2T1B} control samples, and the resulting predictions are compared with the never-blinded A-region data for these samples, which thereby serve as validation samples for the background estimation procedure before it is applied to the high-sensitivity signal region data. In the \texttt{H2T1B} validation sample, the looser requirement on the number of $b$-tagged track-jets associated with the Higgs boson candidate results in the predicted number of multijet events being approximately one order of magnitude larger than the estimated \texttt{H2T2B} signal region background yield. The larger number of events, and the consequentially smaller statistical fluctuations in both the background prediction and the data, renders the \texttt{H2T1B} sample a primary tool for evaluating the performance of the background modelling procedure before it is applied to the signal region.
 
\begin{figure}
\centering
\subfloat[]{{\includegraphics[width=0.44\textwidth]{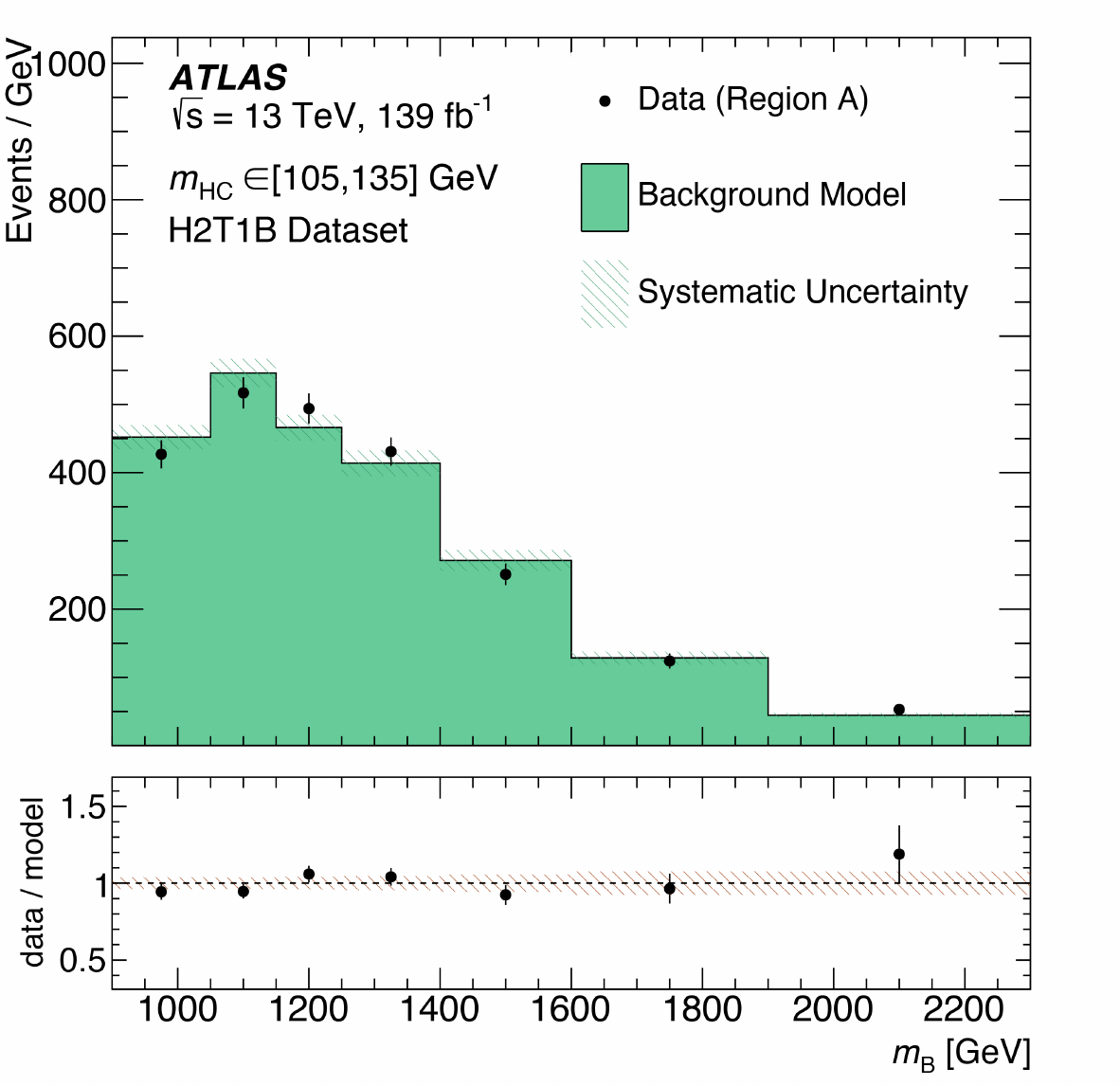}}\label{fig:val_1b}}
\qquad
\subfloat[]{{\includegraphics[width=0.44\textwidth]{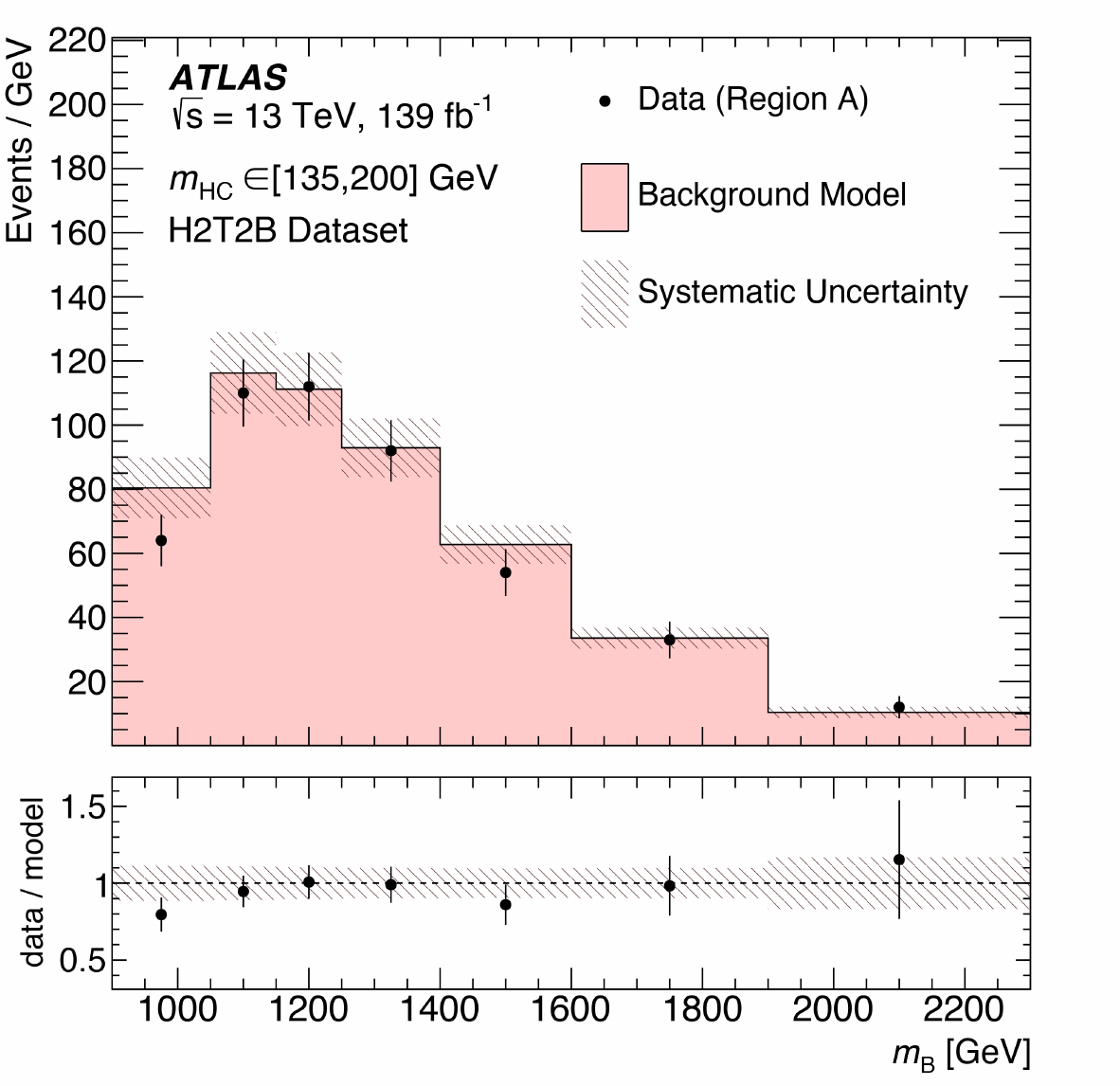}}\label{fig:val_vs}}
\caption{Comparison between ABCD-region A data (points) and predicted background (solid histogram) in (a) the \texttt{H2T1B} data sample and (b) the validation sideband of the \texttt{H2T2B} sample. The pre-fit total systematic uncertainty of each background prediction is indicated by the shaded areas.}\label{fig:validation}
\end{figure}

Figure~\ref{fig:validation} shows the level of agreement between the predicted SM background and the data in the validation samples.  The shaded areas around the background prediction represent the total systematic uncertainty of the background prediction in that sample, defined as the quadrature sum of the individual contributions from the kinematic reweighting of the VLB $\pt$ and small-$R$ jet $\pt$, the $R_\text{corr}$ scaling, and the statistical fluctuations of the B-region data sample, propagated through the modelling procedure (as explained in greater detail in Section~\ref{sec:syst}).
 
In each validation sample  the level of agreement between the SM background prediction and the data is evaluated both visually and via their calculated $\chi^2$ value. No statistically significant or consistently observed discrepancy between data and prediction is detected in any of the validation samples.


\section{Systematic uncertainties}
\label{sec:syst}

 
The expected signal predicted from the simulated signal samples is affected by uncertainties in the modelling of the reconstruction and calibration techniques used in this analysis. These uncertainties do not affect the background prediction, which is entirely derived from data. The systematic uncertainties in the background modelling stem from the uncertainty in the residual-correlation correction $\Delta \rcorr$ and the uncertainty in the procedure used to derive the event weights for the kinematic shape corrections, described in Section~\ref{sec:background}, which are applied to B-region events in order to generate the background prediction for the signal region.
 
\subsection{Signal uncertainties}
 
The uncertainties in the \pt~scale of small-$R$ and large-$R$ jets (JES), as well as the mass scale for large-$R$ jets (JMS), are evaluated by combining information about detector reconstruction performance from studies of MC events and LHC data events~\cite{PERF-2015-04,nowPublic}. The uncertainty due to the jet \pt resolution~\cite{nowPublic} (JER) is obtained from the effect of  smearing the jet \pt in simulated signal events using a $\pt$- and $\eta$-dependent parameterisation of the jet energy resolution derived from \pt-imbalance measurements. The smeared jet $\pt$ distribution is propagated through the event reconstruction and selection, and its effect on the final observable is evaluated. This procedure gives rise to a one-sided systematic uncertainty, which is then symmetrised before performing the statistical fit. The JES and JER uncertainties are modelled in the fit by 30 and 13 nuisance parameters, respectively describing the individual sources of uncertainty for the scale and resolution of the reconstructed jet energy. They are modelled separately for small-$R$ and large-$R$ jets, and nuisance parameters describing the same physical source of uncertainty for the two jet sizes are treated as fully correlated, with the JER uncertainty for large-$R$ jets found to have a negligible effect.
 
The uncertainty due to the large-$R$ jet mass resolution~\cite{ATLAS-CONF-2016-035} (large-$R$ JMR) is obtained by smearing the jet mass using a Gaussian function with a width determined in MC studies of the reconstructed $t$-quark, $W$ boson and Higgs boson mass peaks. The combined impact of all the above uncertainties on the signal yield varies from $6$\% to $13$\%, with the large-$R$ jet mass resolution dominating at both the low end ($6\%$) and high end ($13\%$) of the investigated VLB resonance mass range.
 
The efficiency of flavour tagging in MC events is corrected to match the observed efficiency in data by means of data-derived correction factors calculated at the centres of jet $\pt$ bins and applied to MC events. The uncertainties associated with this correction in the signal Monte Carlo simulation, as well as with their extrapolation in the high \pt region, are estimated by varying the $\eta$- and \pt-dependent correction factors applied to each jet within a range that reflects the systematic uncertainty in the measured $b$-tagging efficiency in data and simulation~\cite{FTAG-2018-01}. The impact of the $b$-tagging uncertainties on the expected signal yield is found to be just under $3\%$ across the VLB mass range of the search. Uncertainties in the $b$-tagging efficiency sharing the same physical source are assumed to be fully correlated across jet collections.

The uncertainty in modelling the $\log{\!\Delta R^*}$ variable is assessed by assuming a conservative track-jet \pt uncertainty, based on the worst case of a single-track track-jet and using the ATLAS track \pt\ resolution, and propagating it through the event selection step that involves $\log{\!\Delta R^*}$. This leads to an asymmetric $+3\%/-6\%$ uncertainty in the expected signal yield. 
A possible uncertainty originating from mismodelling of the showering and hadronisation processes is investigated by comparing the $\log{\!\Delta R^*}$ distribution shape in simulated \PYTHIA[8] and \HERWIG[7] multijet events in the kinematic region targeted by the analysis. No substantial discrepancy is observed, leading to a negligible systematic uncertainty.
 
The impact of a number of theoretical uncertainties on the modelling of signal events is investigated. Both the uncertainty due to missing higher-order corrections in the signal Monte Carlo computation and the uncertainty in the factorisation scale are propagated through the full reconstruction and event selection procedure to assess their impact on the predicted signal yield and distribution of the discriminating variable. To obtain a comprehensive assessment, this procedure was applied to a number of signal resonance mass and width hypotheses covering the theoretical phase space targeted by the search. A conservative 4\% uncertainty in the signal normalisation is introduced to account for the renormalisation scale uncertainty, while the impact of the factorisation scale uncertainty is found to be 5\% across the mass range.
 
The statistical uncertainty of the MC signal simulations originates from the limited size of the simulated samples and is accounted for in the fit by using as many uncorrelated nuisance parameters as there are bins of the final observable, each regulated by Poisson statistics, given the statistical nature of the uncertainty. For each bin, the $\pm 1\sigma$ variation is taken to be the well-known Poisson uncertainty of a weighted event count $\sum_{i,\ i\in \text{bin}}{w_i^2}$, with $w_i$ being the weight assigned to the $i^\text{th}$ event belonging to the bin under consideration.
 
Finally, the uncertainty in the combined 2015--2018 integrated luminosity is 1.7\% \cite{ATLAS-CONF-2019-021}, obtained using the LUCID-2 detector \cite{Avoni_2018} for the primary luminosity measurements.
 
\subsection{Background uncertainties}
The systematic uncertainty in the background prediction comes entirely from the sources of 
uncertainty associated with
the data-driven background estimation procedure outlined in Section~\ref{sec:background}.
 
The per-event kinematic reweighting factors come with an associated uncertainty in the form of $\pm 1\sigma$ weight variations derived from the Kernel regression confidence bands. The reweighting based on the VLB \pt\ produces a relatively constant uncertainty in the background $m_B$ distribution, ranging from $3.5\%$ at the low end of the reconstructed VLB candidate invariant mass range ($m_B=1000$~\GeV) to $4.5\%$ at the high end ($m_B=2000$~\GeV). The second reweighting, derived from the small-$R$ jet \pt\ spectrum, likewise leads to an uncertainty ranging from $\pm5\%$ in the low-mass region to $\pm12\%$ at the high-mass end.
 
A second source of uncertainty, amounting to $\pm 5.1\%$ of the predicted background yield, is associated with the statistical uncertainty of the cross-ratio $\rcorr$ by which the background model is rescaled to compensate for the residual correlation between the two event variables chosen to form the ABCD axes.
 
A third source of uncertainty in the background is the possible dependence of both the per-event kinematic reweighting factors and $k_\text{fwd}$ on the HC mass, which may result in a biased prediction because such quantities are computed globally within the Higgs mass window $m_{\text{HC}} \in [105, 135]$~\GeV. An alternative set of weights and $k_\text{fwd}$ value is extracted from regions C and D of the Higgs mass sideband (\texttt{H2T2B\_VS}) sample and applied to the B-region events in the Higgs mass window (signal region) sample to produce an alternative estimate of the background. The difference between the VS-based prediction and the nominal one is taken as a systematic uncertainty of the reweighting model's stability with respect to the HC mass. The predicted impact on the predicted SM background yield ranges from below $1\%$ for the lowest values of the probed mass range to about $4\%$ above $2~\TeV$.
 
Finally, the shape of the predicted background is affected by the bin-wise Poisson fluctuations in the B-region data used to construct the model. 
The uncertainty in each $m_B$ bin of the final model originating from this source is computed by propagating the Poisson uncertainty of the relevant bin through the whole modelling procedure, and accounts for an uncertainty in the predicted yields ranging from $7\%{-}8\%$ for $m_{B} \in [850, 1450]~\GeV$ to $10\%{-}20\%$ for $m_{B} \in [1450, 2250]~\GeV$, where it is the dominant uncertainty in the background prediction.
 
Table~\ref{tab:syst_full} summarises the impact of each category of systematic uncertainties on the yield of each of the affected event samples. To highlight the changing impact and hierarchy of the systematic uncertainty sources across the investigated resonance mass range,  relative systematic uncertainties are displayed for both a relatively low-mass $1.2~\TeV$ resonance and the highest-mass simulated sample ($2.0~\TeV$). The two single-VLB production modes are examined separately and denoted by ZBHb ($bZ\rightarrow B$ vertex) and WBHb ($tW \rightarrow B$ vertex).

\begin{table}[htbp]
\centering
\caption{Relative effect of the pre-fit systematic uncertainties from each group of sources on the yields of the predicted background and two simulated signals with $\kappa = 0.5$ and VLB mass equal to 1.2 or 2.0~\TeV. The $W$-initiated (WBHb) and $Z$-initiated (ZBHb) production modes are kept separate. The size of the systematic uncertainties affecting the background is similarly provided for reconstructed VLB candidates separately for 1.2~\TeV and 2.0~\TeV.}
\label{tab:syst_full}
\begin{tabular}{c|c|c|c|c||c|c}
& \multicolumn{2}{c|}{\textbf{VLB mass = 1.2 TeV}} &  \multicolumn{2}{c||}{\textbf{VLB mass = 2.0 TeV}} & \textbf{At 1.2 TeV} & \textbf{At 2.0 TeV} \Tstrut\\[0.5ex]
\hline
\textbf{Systematic} & \textbf{ZBHb} & \textbf{WBHb} & \textbf{ZBHb} & \textbf{WBHb} & \multicolumn{2}{c}{\textbf{Background}} \TBstrut\\[0.5ex]
\hline
\hline
\textit{b}-tagging & 2.8\% & 2.8\% & 2.9\% & 2.8\% & \multicolumn{2}{c}{\slashbox{}{}}\TBstrut\\[0.5ex]
 
JER & 3.4\% & 1.8\% & 2.3\%& 4.6\% &\multicolumn{2}{c}{\slashbox{}{}}\Tstrut\\[0.5ex]
JES & 4.4\% & 2.1\% & 2.9\% & 2.0\%& \multicolumn{2}{c}{\slashbox{}{}}\Tstrut\\[0.5ex]
 
Large-$R$ JES & 1.9\% & 0.4\% & 5.4\% & 3.1\% &\multicolumn{2}{c}{\slashbox{}{}}\Tstrut\\[0.5ex]
Large-$R$ JMR & 6.1\% &10.5\%\phantom{0} & 12.0\%\phantom{0} &13.0\%\phantom{0} &\multicolumn{2}{c}{\slashbox{}{}}\Tstrut\\[0.5ex]
Large-$R$ JMS & 2.5\% & 4.3\% & 1.4\% & 2.7\% &\multicolumn{2}{c}{\slashbox{}{}}\Tstrut\\[0.5ex]
$\log{\!\Delta R^*}$ &  \multicolumn{4}{c}{+3\% / $-6\%$}&\multicolumn{2}{c}{\slashbox{}{}}\Tstrut\\[0.5ex]
 
Luminosity & 1.7\% & 1.7\% & 1.7\% & 1.7\% &\multicolumn{2}{c}{\slashbox{}{}}\Tstrut\\[0.5ex]
MC statistics & 5.1\%& 5.6\% &5.9\% &5.6\% & \multicolumn{2}{c}{\slashbox{}{}}\Tstrut\\[0.5ex]
Renormalisation scale & \multicolumn{4}{c||}{4\%} & \multicolumn{2}{c}{\slashbox{}{}}\Tstrut\\[0.5ex]
Factorisation scale & \multicolumn{4}{c||}{5\%} & \multicolumn{2}{c}{\slashbox{}{}}\Tstrut\\[0.5ex]
VLB \pt weights &\slashbox{}{}&\slashbox{}{}&\slashbox{}{}&\slashbox{}{}& \phantom{0.0}3.5\% & \phantom{0.0}4.5\%\Tstrut\\[0.5ex]
$R=0.4$ jet \pt weights &\slashbox{}{}&\slashbox{}{}&\slashbox{}{}&\slashbox{}{}& \phantom{0}5\% & 12\%\Tstrut\\[0.5ex]
Model stability &\slashbox{}{}&\slashbox{}{}&\slashbox{}{}&\slashbox{}{}& \phantom{0.0}0.8\% & \phantom{0}4\% \Tstrut\\[0.5ex]
$R_\text{corr}$ &\slashbox{}{}&\slashbox{}{}&\slashbox{}{}&\slashbox{}{}& \multicolumn{2}{c}{5.1\%} \Tstrut\\[0.5ex]
B-region statistics &\slashbox{}{}&\slashbox{}{}&\slashbox{}{}&\slashbox{}{}& \phantom{0}8\% & 18\% \Tstrut\\[0.5ex]
\hline
\textbf{TOTAL} & \textbf{13\%} & \textbf{14\%}& \textbf{17\%}&\textbf{15\%} &\textbf{11\%} & \textbf{23\%} \TBstrut\\[0.5ex]
\end{tabular}

\end{table}


\section{Statistical analysis}\label{sec:stats}
The statistical interpretation of the data is carried out using a binned maximum-likelihood fit to the invariant mass ($m_B$) distribution of the reconstructed VLB candidates, based on the expected signal and background yields. The likelihood model is defined as:
 
\begin{equation}\nonumber{\cal L} = \prod_i P_{\text{pois}}(n_i|\lambda_i) \times {\cal N}({\bm \theta})
\end{equation}
 
where $P_{\text{pois}}(n_i|\lambda_i)$ is the Poisson probability to observe $n_i$ events when $\lambda_i$ events are expected in bin $i$ of the $m_B$ distribution, and ${\cal N}({\bm \theta})$ is a series of Gaussian or log-normal distributions for the nuisance parameters, ${\bm \theta}$, corresponding to the systematic uncertainties related to the signal and background yields in each bin. The $\lambda_i$ are expressed as $\lambda_{i} = \mu s_{i} ({\bm \theta}) + b_{i}({\bm \theta})$, with $\mu$ being the signal strength, defined as a signal cross-section in units of the theoretical prediction, left as the free-floating parameter of interest to be determined by the fit, and $s_{i}( {\bm \theta })$ and $b_{i}( {\bm \theta })$ being the expected numbers of signal and background events, respectively. Nuisance parameters are allowed to float in the fit within their constraints ${\cal N}({\bm \theta})$, and thus alter the normalisation and shape of the signal $m_B$ as well as the background $m_B$ distribution.
 
Signal spillage outside the signal region (as introduced in Section~\ref{sec:background}) has a potential effect on the background model. This is compensated for by subtracting the signal contribution in region B from the background at the fitting stage.
Before subtraction, region-B signal events are scaled by $k_\text{fwd}\times R_\text{corr}$ and kinematically reweighted using the same weight functions employed in the background modelling phase, in order to reproduce their actual contribution to the signal region background model.
The sample subtraction is effectively accomplished by assigning a negative signal-strength parameter $\mu_\text{cont}$ to the contamination sample and constraining it to the fitted value of the parameter of interest $\mu$ by setting $\mu_\text{cont} = -\mu$. This operation ensures that both the signal yield and the magnitude of the resulting background contamination are simultaneously determined in the statistical fit.
 
As with the principal signal MC  simulations, the subtracted contamination sample is affected by systematic uncertainties related to the theoretical and experimental understanding of single-VLB production and its detection at ATLAS. Consequently, the nuisance parameters describing such uncertainties are set to be fully correlated across all MC-derived samples.
In addition, the contamination sample is also affected by the uncertainties introduced by the kinematic reweighting procedure applied to the region-B signal simulation in order to estimate the effect of signal spillage on the background model. The nuisance parameters describing the impact of such uncertainties in the fit are set to be fully correlated with those for the same uncertainty source in the QCD multijet background model.
 
Information about the signal strength $\mu$ is extracted from a signal-plus-background likelihood fit to the data, using a test statistic based on the profile likelihood ratio. The distributions of the test statistic under the signal-plus-background and background-only hypotheses are obtained using asymptotic formulae~\cite{asymptotics}. The systematic uncertainties with the largest post-fit impact on $\mu$ at $m_B=1.2$~\TeV\ and $m_B=2.0$~\TeV\ are the uncertainty in the $R_\text{corr}$ correction and the uncertainty in the reweighting of the small-$R$ jet $p_{\mathrm{T}}$, respectively. 
The level of agreement between the observed data and the background prediction is assessed by computing the local \textit{p}-value $p_0$ for the observed value of the \textit{profile likelihood} test statistic given its asymptotic distribution for the background-only hypothesis. The value of $p_0$ is defined as the probability to observe an excess at least as large as the one observed in data, under the background-only hypothesis. 
Expected and observed upper limits are set at 95\% confidence level (CL) on the cross-section for single-VLB production in the decay channel under investigation ($\sigma(pp \rightarrow B \rightarrow bH)$)  using the CL$_\text{s}$ prescription~\cite{CLs_2002}.
 
The limit-setting procedure is iterated over all available simulated signal hypotheses, defined by the resonance mass and the value of the coupling strength $\kappa$ (the parameter regulating the resonance width, as introduced in Section~\ref{sec:samples}), allowing a broad experimental exploration of the phenomenological phase space of single-VLQ production.

\section{Results}
 
\begin{figure}[h]
\centering
\includegraphics[width=0.60\linewidth]{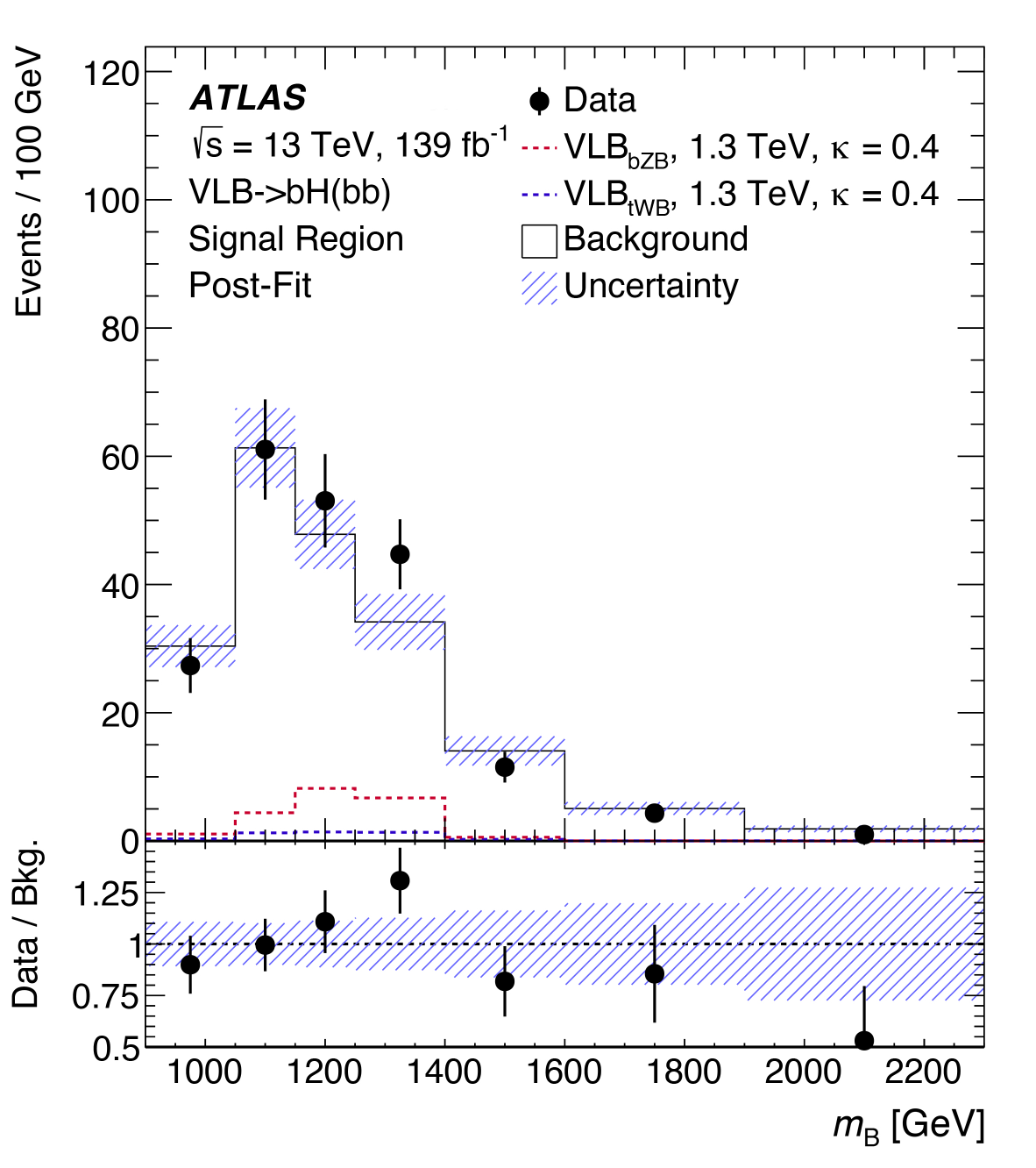}
\caption{Comparison between data (black dots) and the SM data-derived background model (solid white) after a binned maximum-likelihood background-only fit in the signal region. The shaded area represents the total systematic uncertainty of the background model. The expected contribution of a $m_B = 1.3$~\TeV, $\kappa = 0.4$ signal produced via either the $Z$-initiated (red dotted, labelled $\text{VLB}_\text{bZB}$) or $W$-initiated (blue-dotted) mode is overlaid for reference. The first and last bins include underflow and overflow events. The lower panel shows the bin-wise ratios of the data and background yields.}
\label{fig:prepost_bonly}
\end{figure}

Following the statistical interpretation scheme outlined in Section~\ref{sec:stats}, a binned maximum-likelihood fit is performed on the signal region data. The data yield in the signal region, and the background yields and their relative uncertainties before and after the statistical fit, are shown in Table~\ref{tab:prepost_bonly}. The $m_{B}$ spectrum in data is compared with the post-fit background model in Figure~\ref{fig:prepost_bonly}. The expected event distribution for a 1.3~\TeV\  isospin-singlet VLB with $\kappa = 0.4$ is overlaid on the data distribution, with the  $Z$-initiated and $W$-initiated contributions displayed separately. The $Z$-initiated production mode dominates the predicted signal region yields (approximately 80\%). For a $(B,Y)$ doublet, only the $Z$-initiated production mode contributes, with a larger signal yield expected due to the larger $\xi_Z$ value as discussed in Section~\ref{sec:intro}.

Since the data and predicted background generally agree well, the maximum-likelihood fit does not noticeably shift the nuisance parameters affecting the background model from their nominal values.

\label{sec:result}
\begin{table}[h]
\centering
\caption{Comparison of the yields of the SM background and a VLB singlet, for various mass and $\kappa$ values, before and after a background-only fit. The post-fit signal yield is accordingly zero by definition (conditional fit with $\mu=0$). The observed data yield in the signal region is also displayed.}
\label{tab:prepost_bonly}
\begin{tabular}{c||c||c|c}
\multicolumn{2}{c||}{\textbf{Sample}} &  \textbf{Pre-Fit} & \textbf{Post-Fit}\Tstrut\\[0.5ex]
\hline
\multirow{3}{*}{VLB, 1.2 TeV} & $\kappa=0.4$& $33 \pm 6$ & \slashbox{}{} \TBstrut\\[0.5ex]
& $\kappa=0.5$& \phantom{0}$51 \pm 10$ & \slashbox{}{} \TBstrut\\[0.5ex]
& $\kappa=0.6$& \phantom{0}$66 \pm 13$ & \slashbox{}{} \TBstrut\\[0.5ex]
\hline
\multirow{3}{*}{VLB, 1.6 TeV} & $\kappa=0.4$& $21 \pm 4$ & \slashbox{}{} \TBstrut\\[0.5ex]
& $\kappa=0.5$& $34 \pm 6$ & \slashbox{}{} \TBstrut\\[0.5ex]
& $\kappa=0.8$& \phantom{0}$50 \pm 10$ & \slashbox{}{} \TBstrut\\[0.5ex]
\hline
\multirow{3}{*}{VLB, 2.0 TeV} & $\kappa=0.4$& \phantom{0}$9 \pm 3$ & \slashbox{}{} \TBstrut\\[0.5ex]
& $\kappa=0.5$& $16 \pm 4$ & \slashbox{}{} \TBstrut\\[0.5ex]
& $\kappa=0.8$& $23 \pm 6$ & \slashbox{}{} \TBstrut\\[0.5ex]
\hline
\multicolumn{2}{c||}{Background estimate} & $257 \pm 25$ & $260\pm17$ \Tstrut\\[0.5ex]
\hline
\hline
\multicolumn{2}{c||}{\textbf{Data}} & \multicolumn{2}{c}{\textbf{262}}\Tstrut\\[0.5ex]
\end{tabular}
\end{table}
 
\begin{figure}[h]
\centering
\subfloat[]{\includegraphics[width=0.44\textwidth]{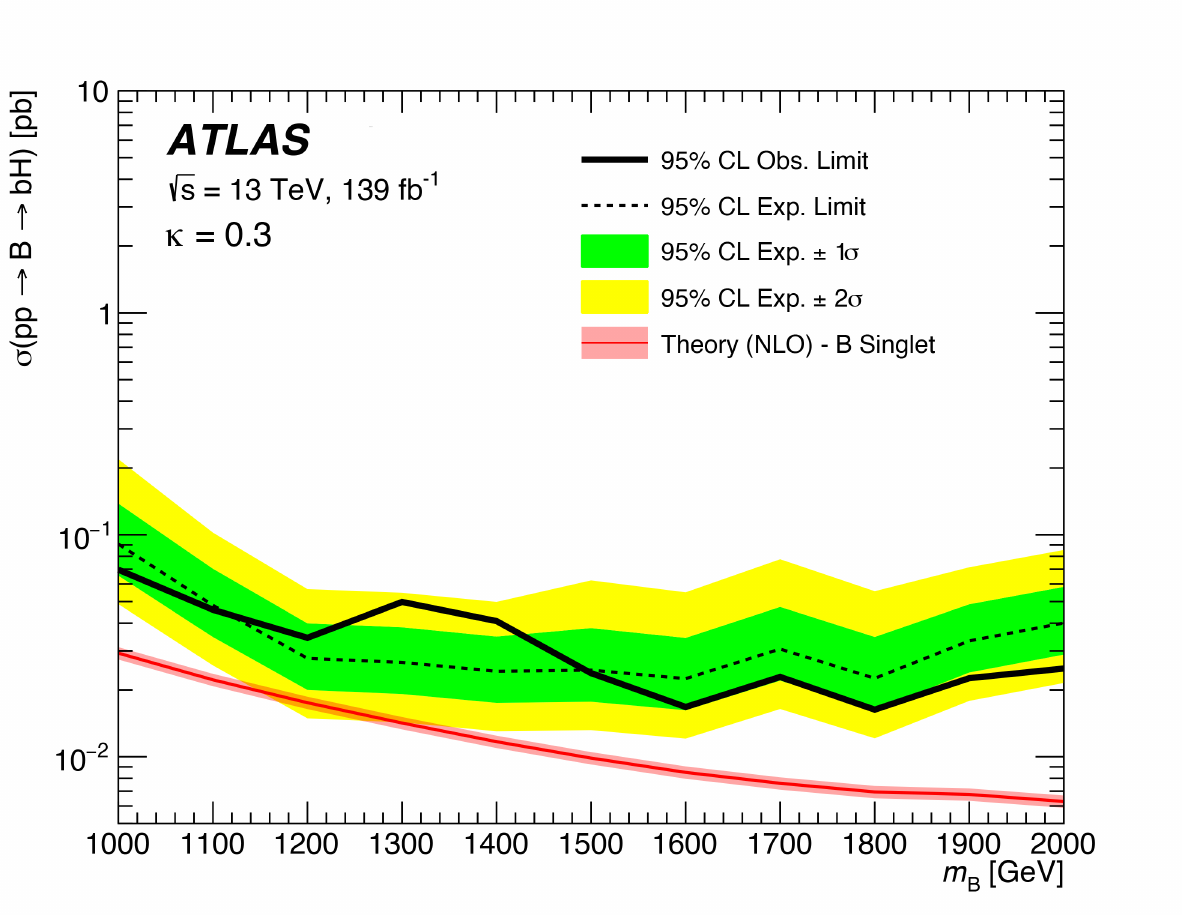}\label{fig:1d_sing03}}
\subfloat[]{\includegraphics[width=0.44\textwidth]{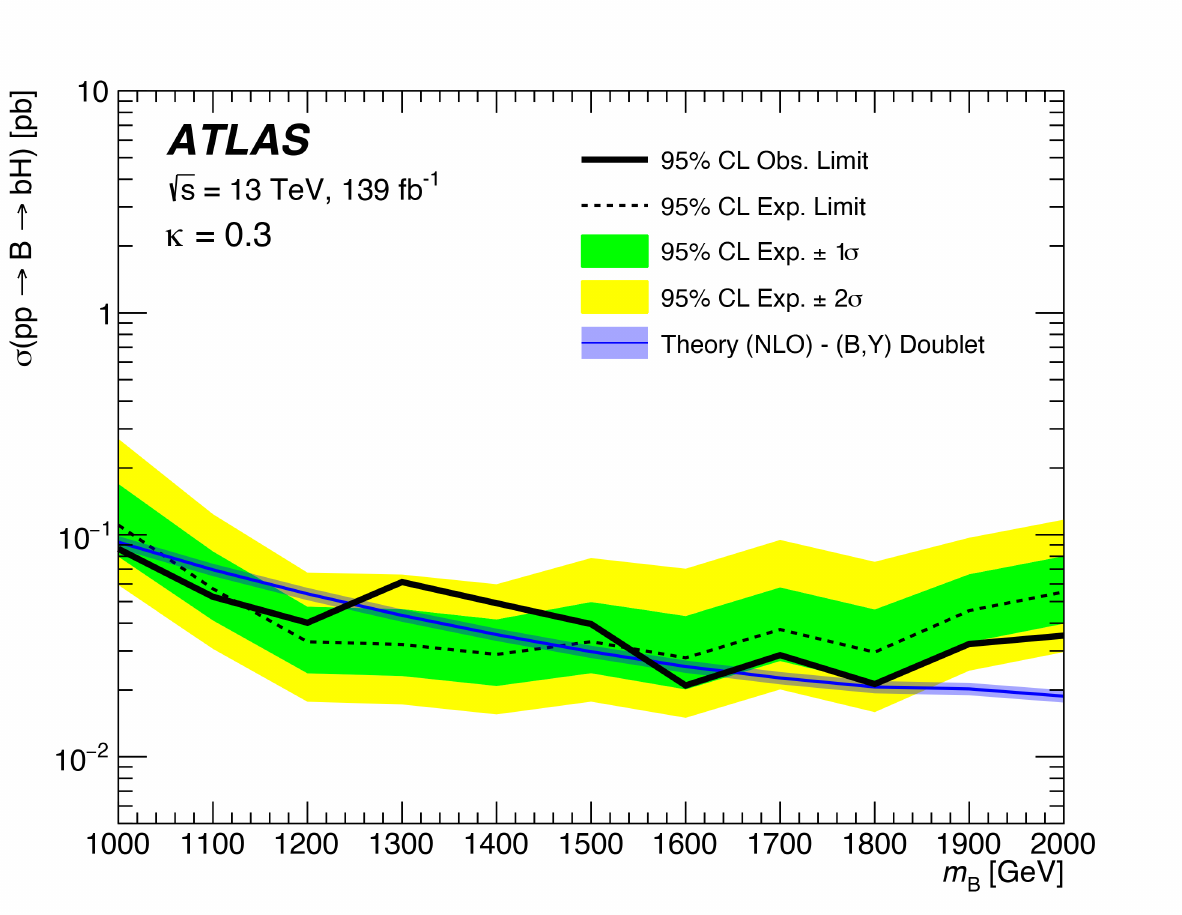}\label{fig:1d_doub03}}
 
\subfloat[]{\includegraphics[width=0.44\textwidth]{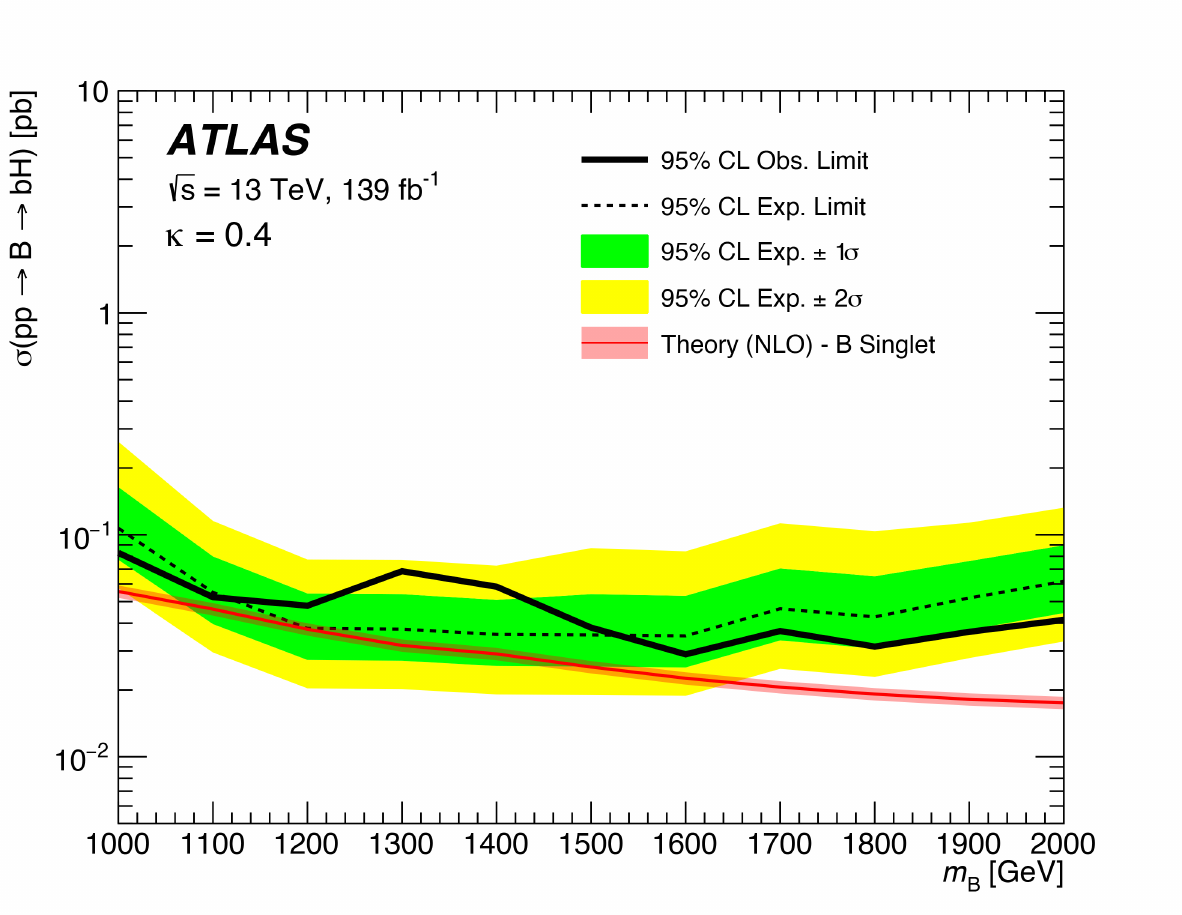}\label{fig:1d_sing04}}
\subfloat[]{\includegraphics[width=0.44\textwidth]{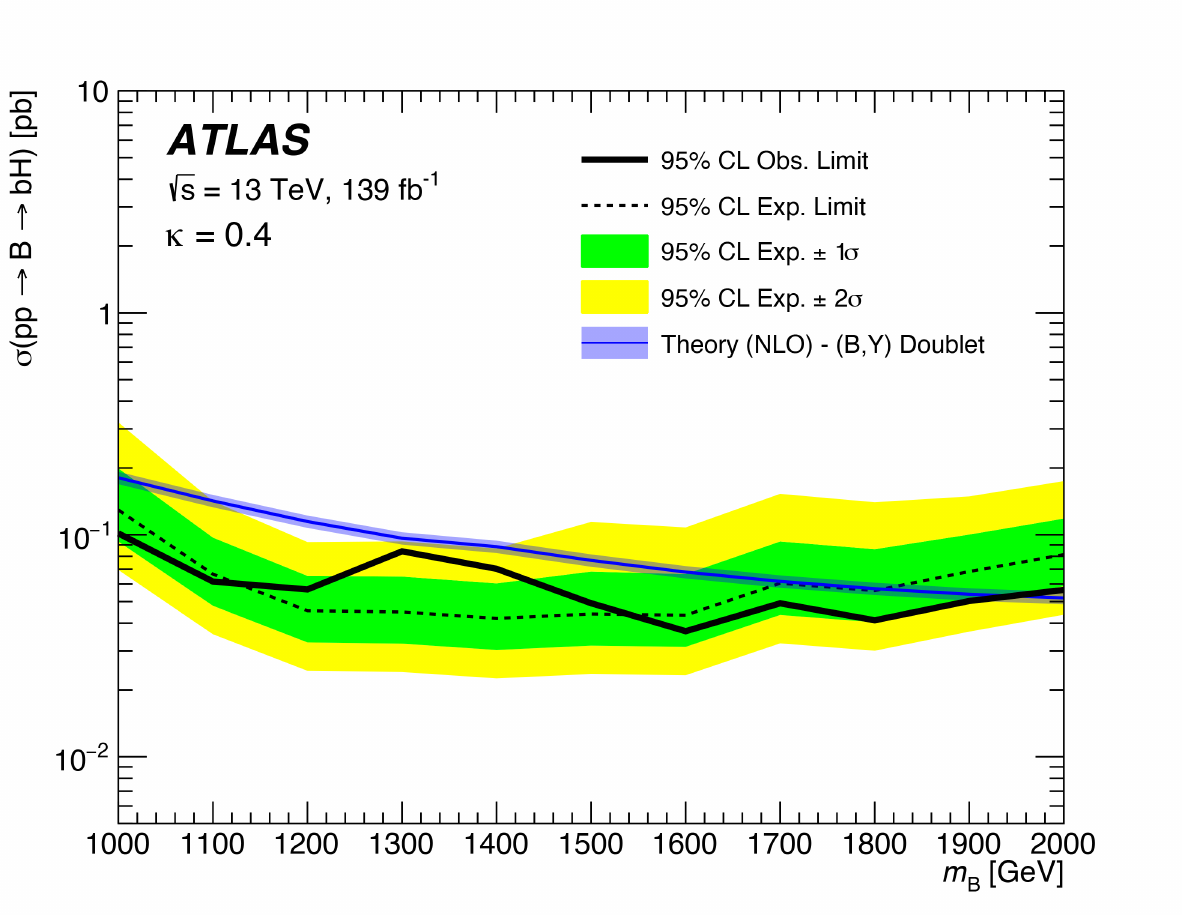}\label{fig:1d_doub04}}
 
\subfloat[]{\includegraphics[width=0.44\textwidth]{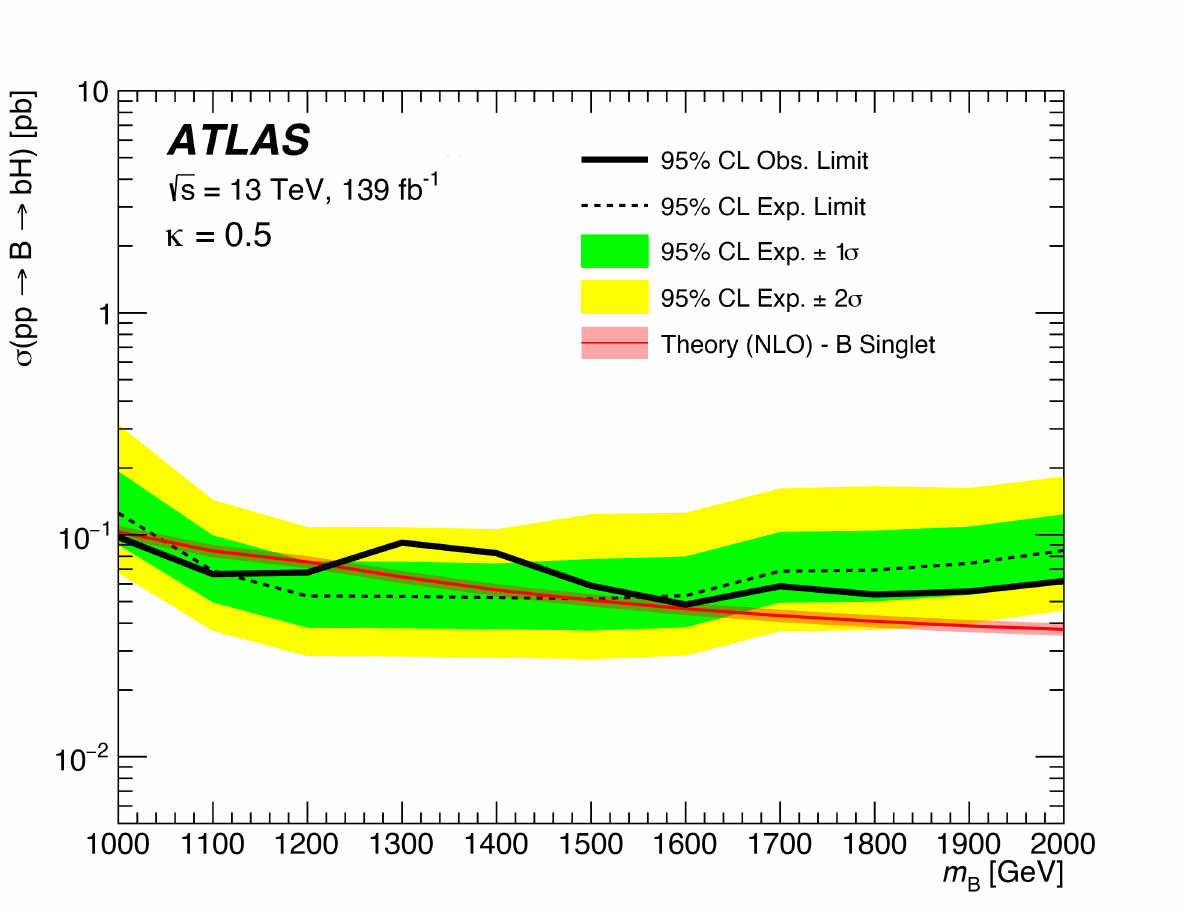}\label{fig:1d_sing03}}
\subfloat[]{\includegraphics[width=0.44\textwidth]{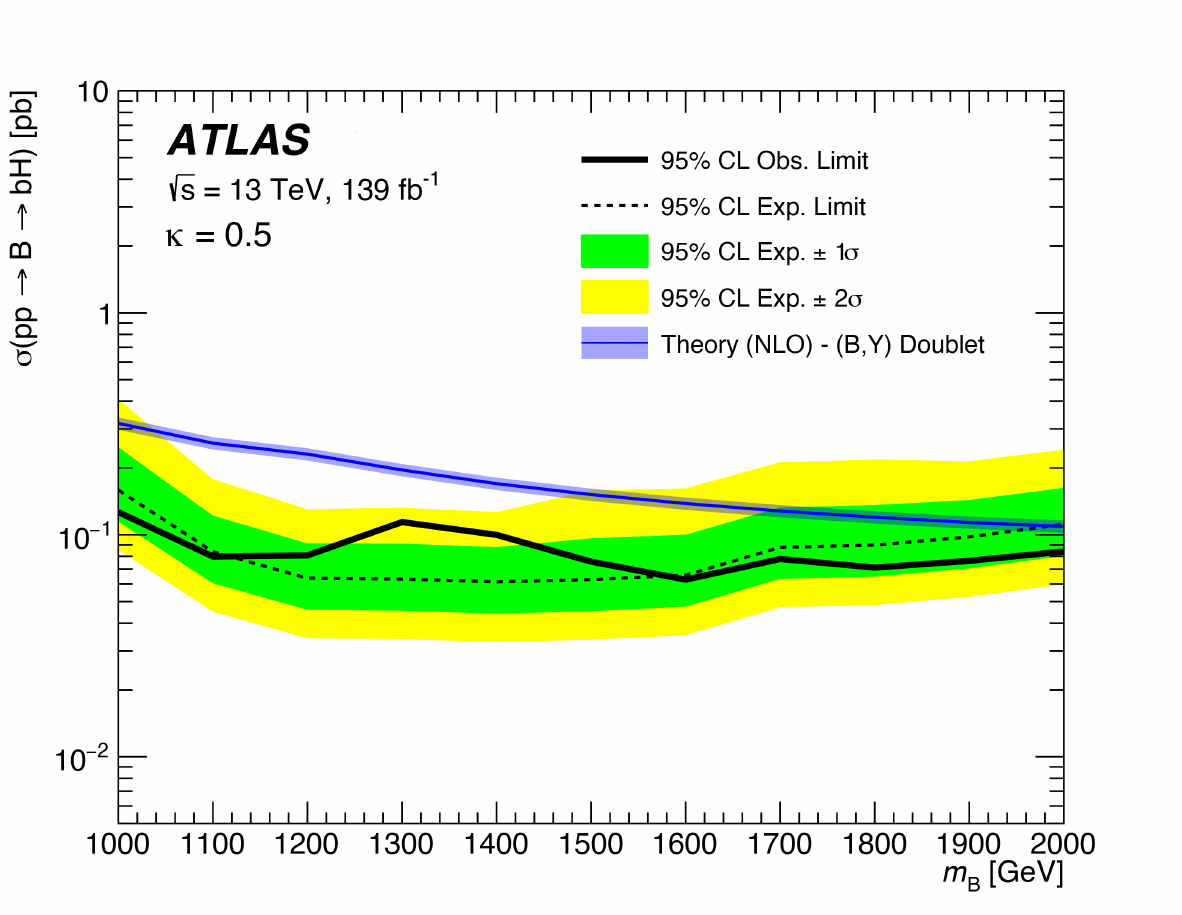}\label{fig:1d_doub04}}
 
\caption{Expected (dashed line) and observed (solid black line) $95\%$ CL exclusion limits on the cross-section for single-VLB  production and the Higgs decay mode in the $B$ singlet (left column) and $(B,Y)$ doublet (right column) theoretical scenarios.   The green and yellow bands about the expected limits represent the $\pm 1\sigma$ and $\pm 2\sigma$ confidence intervals of the expected limits, respectively. The red and blue solid lines and shaded areas respectively trace the evolution of the singlet and doublet production cross-sections as a function of the resonance mass, and their relative theoretical uncertainty accounts for uncertainties in the renormalisation and factorisation scales. Limits are presented for three values of the coupling strength $\kappa$: 0.3 (a,b), 0.4 (c,d) and  0.5 (e,f).}
\label{fig:1d_limits}
\end{figure}

No significant excess of data over the SM background prediction is observed. The largest discrepancy between data and the Standard Model background contribution is observed between 1.3~\TeV\ and 1.4~\TeV, corresponding to a local $p_0 = 0.06$ for the $m_B=1.4$~\TeV, $\kappa = 0.4$ signal hypothesis. The \textit{profile likelihood} technique introduced in Section~\ref{sec:stats} is used to set $95\%$ CL exclusion limits on the production cross-section for a number of theoretical benchmark scenarios defined by the assumed coupling strength $\kappa$ and isospin multiplet state of the VLB.
Figure~\ref{fig:1d_limits} displays mass-dependent $95\%$ CL exclusion limits on the production cross-section for three different values of the coupling strength $\kappa$. The blue and red solid lines overlaid on the exclusion plots represent the calculated NLO single-production cross-section for a VLB occurring in a $(B,Y)$ isospin doublet or an isospin singlet, respectively. The phenomenological properties of the VLB resonance in the two scenarios are inferred by setting $\xi_W = 0.5$, $\xi_Z=\xi_H = 0.25$ for a VLB singlet and $\xi_W = 0$, $\xi_Z=\xi_H=0.5$ for the $(B,Y)$ doublet. The values of the $\xi$ couplings most notably affect the branching fractions of the VLB decay and which production vertex is dominant in single-production processes. For instance, a VLB occurring as part of a $(B,Y)$ doublet can only be produced via the $Z$-initiated vertex because $\xi_W = 0$ in that scenario.

The resonance mass ranges excluded at $95\%$ CL for a set value of $\kappa$ and a specific multiplet state are inferred from Figure~\ref{fig:1d_limits} as the mass intervals for which the expected or observed upper limit on the cross-section is smaller than the relevant theoretical prediction.
A noticeable feature of the $\kappa$-specific limits shown in Figure~\ref{fig:1d_limits} is the approximately constant exclusion power of the search, in terms of production cross-section, for $m_{B} > 1.4~\TeV$. This is understood as being a consequence of the growing dominance of non-resonant $t$-channel production in large-width scenarios,\footnote{Since the total width is well approximated as $\Gamma_B \sim m_B^3\kappa^2$, it is a cubic function of the $B$ mass for constant values of $\kappa$.} which greatly reduces the ability to distinguish states of different mass for $m_B \geq 1.5~\TeV$.
 
\begin{figure}[h]
\centering
\subfloat[]{\includegraphics[width=0.48\linewidth]{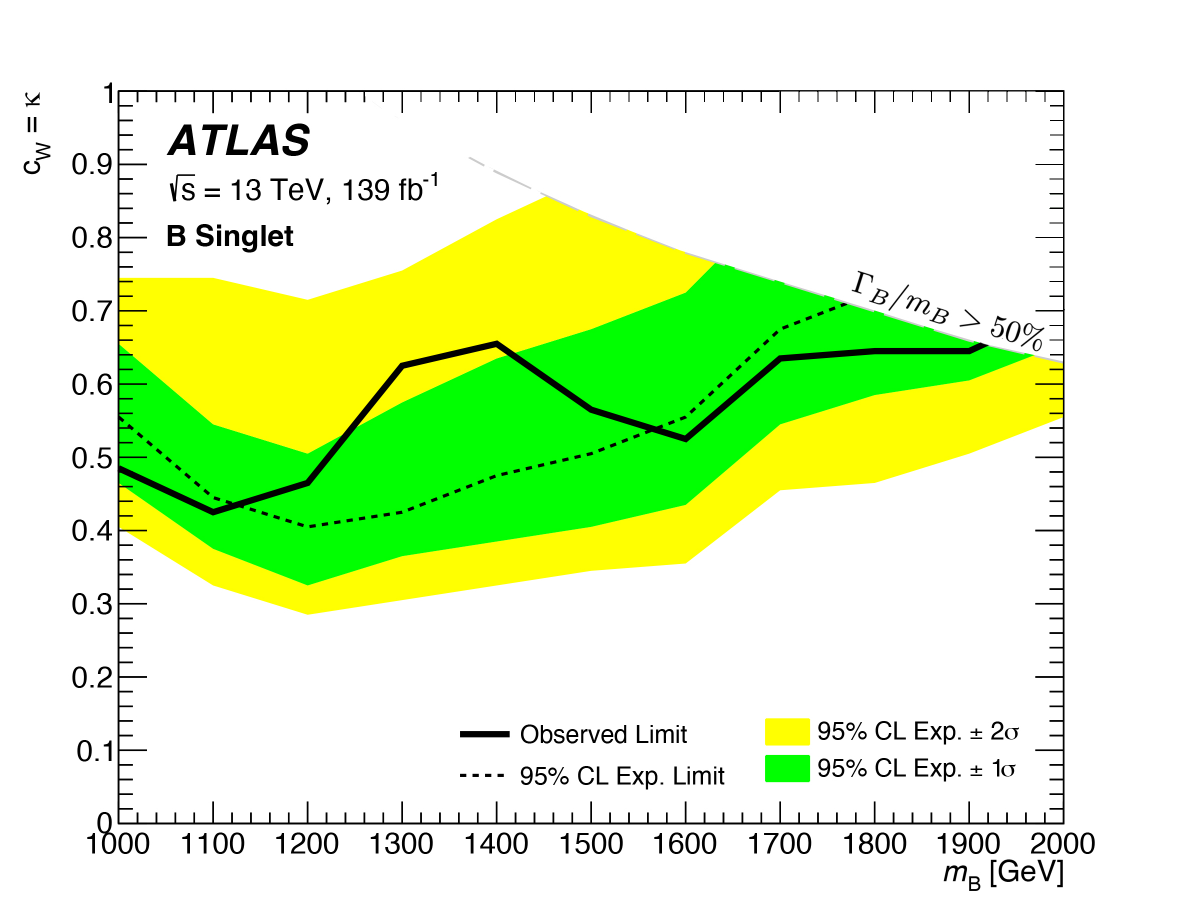}\label{fig:2d_singlet}}
\subfloat[]{\includegraphics[width=0.48\linewidth]{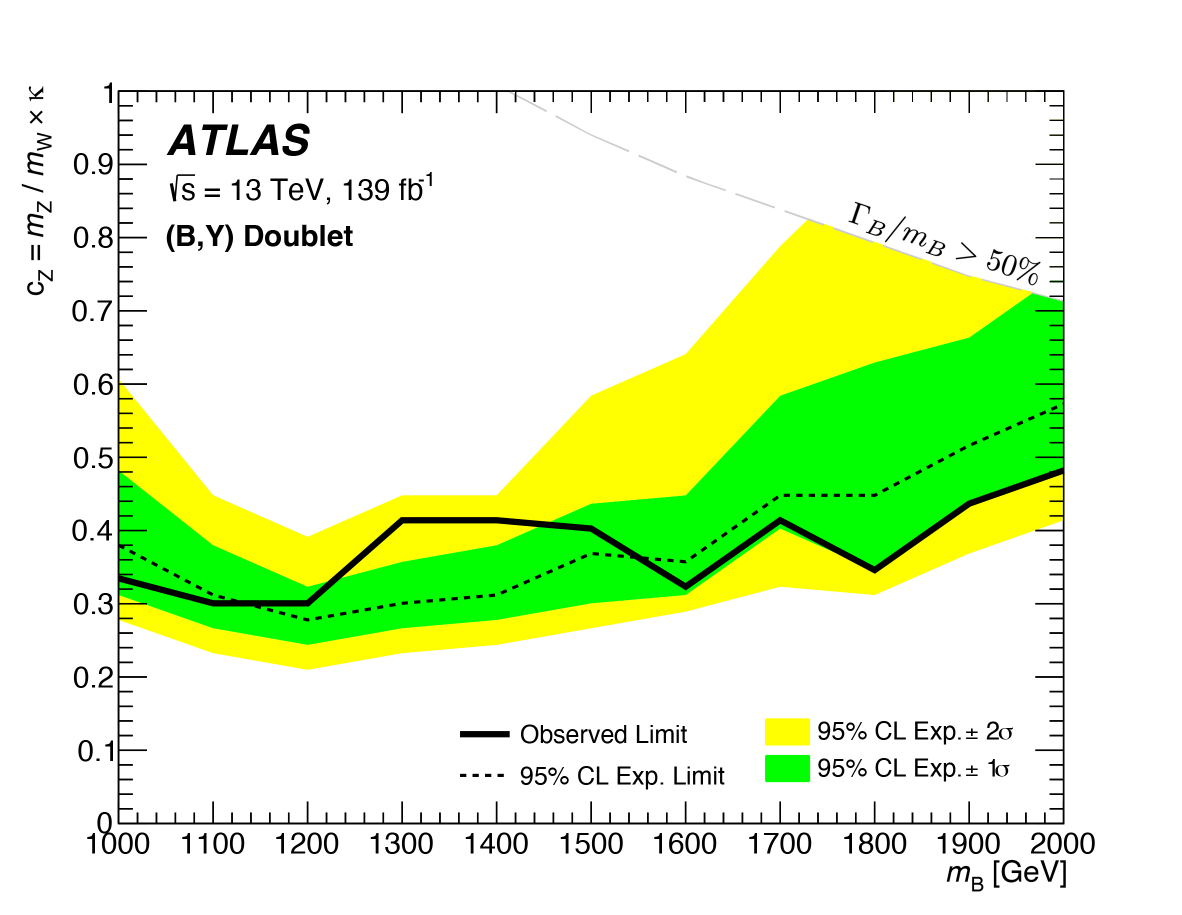}\label{fig:2d_doublet}}\\
\caption{Mass-dependent expected (dashed line) and observed (solid line) $95\%$ CL exclusion limits on the $c_W$ and $c_Z$ phenomenological couplings in the (a) VLB singlet and (b) $(B,Y)$ doublet scenarios, respectively. Results are only displayed for ($m_{B}$, $\kappa$) configurations yielding a fractional resonance width no greater than 50\%. }
\label{fig:2d_coupling}
\end{figure}

The cross-section limits presented above are then interpreted in the form of mass-dependent $95\%$ CL exclusion limits on the value of any of the three phenomenological Lagrangian couplings $c_W$, $c_Z$, and $c_H$, all of which scale with the coupling strength $\kappa$ within a specific multiplet state, as introduced in Eq.~(\ref{eqn:couplings}). Figure~\ref{fig:2d_coupling} displays the expected and observed limits on $c_W$ for a VLB quark occurring as an isospin singlet, and on $c_Z$ in the case of a $(B,Y)$ doublet.
Limits on $c_W$ are derived by examining the production cross-section exclusion limit observed for each available signal benchmark, arranged on a $(m_B,\ \kappa)$ grid. For each available resonance mass, the lowest value of $\kappa$ for which the signal is excluded is converted into a value of $c_W$ through the $c_W = \kappa$ identity for a VLB singlet. The expected and observed limits are only displayed for configurations in the ($m_B$, $\kappa$) space corresponding to values of the fractional resonance width $\Gamma_B/m_B < 50\%$, in accordance with the current practice for interpreting ATLAS results in terms of singly produced VLQs.
 
\begin{figure}[h]
\centering
 
\subfloat[]{\includegraphics[width=0.76\linewidth]{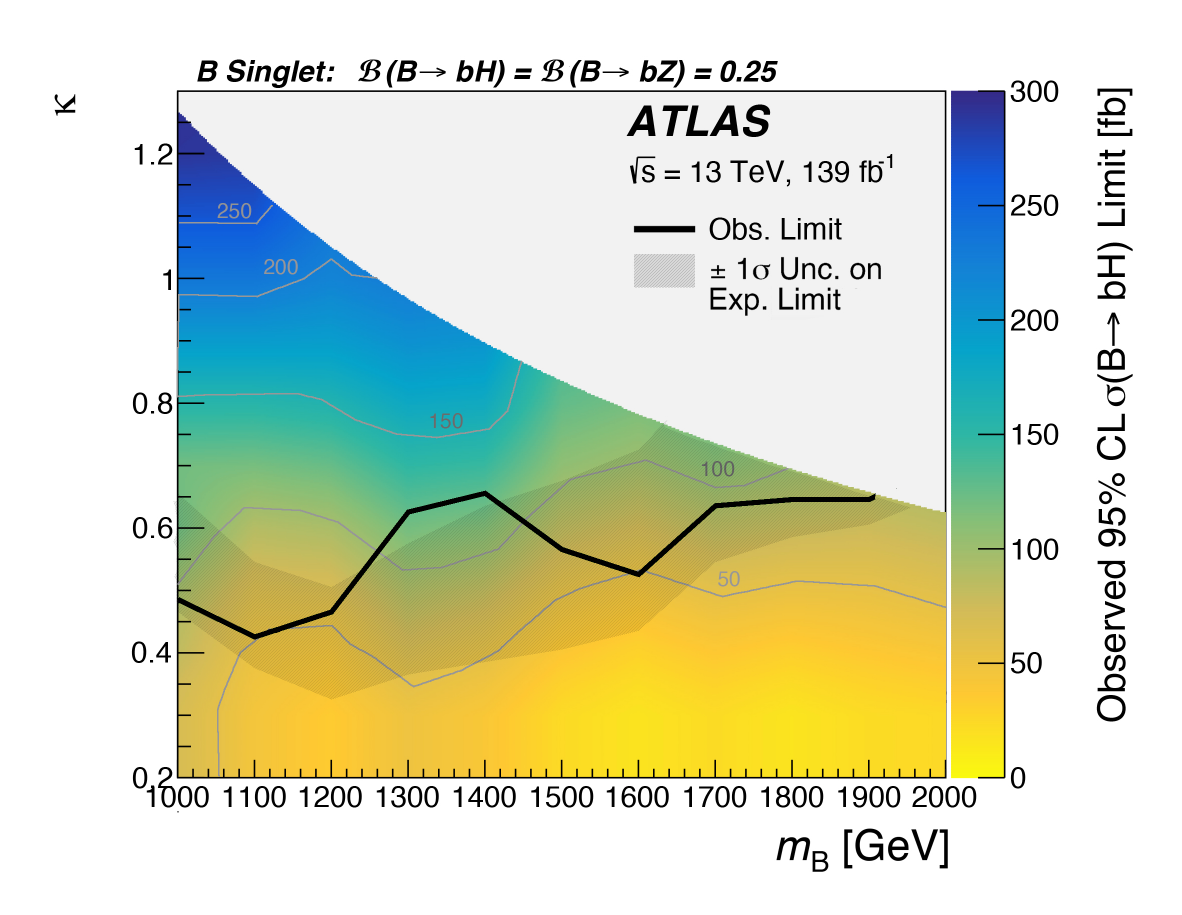}\label{fig:newK_singlet}}
 
\subfloat[]{\includegraphics[width=0.76\linewidth]{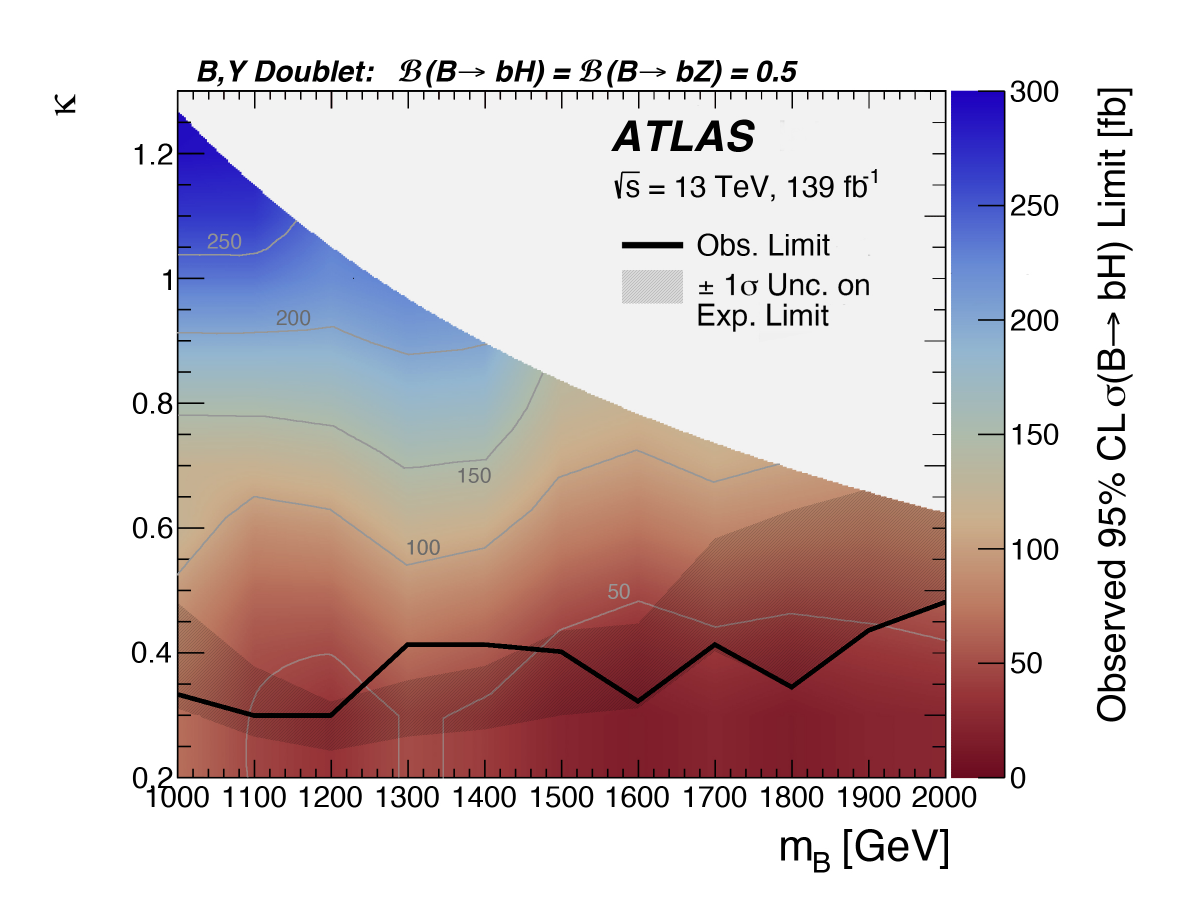}\label{fig:newK_doublet}}\\
 
\caption{Observed limits on the single-production cross-section ($z$-axis) as a function of the resonance mass and coupling strength $\kappa$ in the (a) singlet and (b) doublet scenarios.
To guide the eye, labelled contours are overlaid to identify configurations where the excluded cross-section is an exact multiple of $50~\text{fb}$.
The area above the black overlaid line is the excluded ($m_B, \kappa$) phase-space region, where the experimentally excluded cross-section is greater than the theoretical value. The darker shaded area overlapping the observed exclusion contour indicates the $1\sigma$ confidence band for the expected position (not shown) of the exclusion contour.
}
\label{fig:2d_width}
\end{figure}

The different choice of coupling on the $y$-axis for the doublet limits (Figure~\ref{fig:2d_doublet}) is dictated by the fact that $\xi_W = 0$ for the $(B,Y)$ state implies  $c_W = 0$. Nevertheless, the exclusion power in the singlet and doublet scenarios can be compared as $c_W \simeq \kappa$ in the singlet state, and $c_Z \simeq m_Z/m_W \times \kappa$ for the doublet. The larger predicted cross-section for the $(B,Y)$ doublet than for the singlet results in a larger portion of the mass--coupling phase space being excluded. As an example, the data excludes $c_W > 0.4$, corresponding to $\kappa > 0.4$ for a VLB singlet with $m_B = 1.2~\TeV$, while a VLB of the same mass belonging to a $(B,Y)$ doublet is excluded for $c_Z > 0.20$, corresponding to approximately $\kappa > 0.17$.
Figure~\ref{fig:2d_width} shows the expected and observed exclusion limits on $\sigma(pp\rightarrow B \rightarrow bH)$ as a function of the resonance mass and relative width in the isospin-singlet and isospin-doublet scenarios.
 
Lastly, the results are interpreted beyond the established singlet and doublet scenarios discussed so far. This is accomplished by expressing the experimental limits in a way applicable to the broader and more generic set of configurations in the three-dimensional theoretical phase space defined by the parameter vector $(\xi_W,\ \xi_Z,\ \xi_H,\ \kappa)$ and the constraint $\sum{\xi_i} = 1$.
For the sake of presentation, the three-dimensional parameter space is collapsed into two independent quantities by imposing the theoretically well-motivated \cite{Aguilar-Saavedra:2013qpa} constraint $\xi_Z = \xi_H$, which applies to every multiplet state envisioned in the primary theoretical models for VLQs.
 
Given this premise, the resulting set of independent parameters $(\xi_W,\ \kappa)$ can be rotated to provide a more phenomenology-oriented visualisation of the search results by means of the following transformation:
 
\begin{equation}
\nonumber
(\xi_W,\ \kappa)\longrightarrow(\xi_W,\ m_B^2\kappa^2)\propto{\Bigg(\mathcal{B}(B\rightarrow tW),\  \frac{\Gamma_B}{m_B}\Bigg)}.
\end{equation}
 
\begin{figure}[h]
\centering
\subfloat[]{{\includegraphics[width=0.76\textwidth]{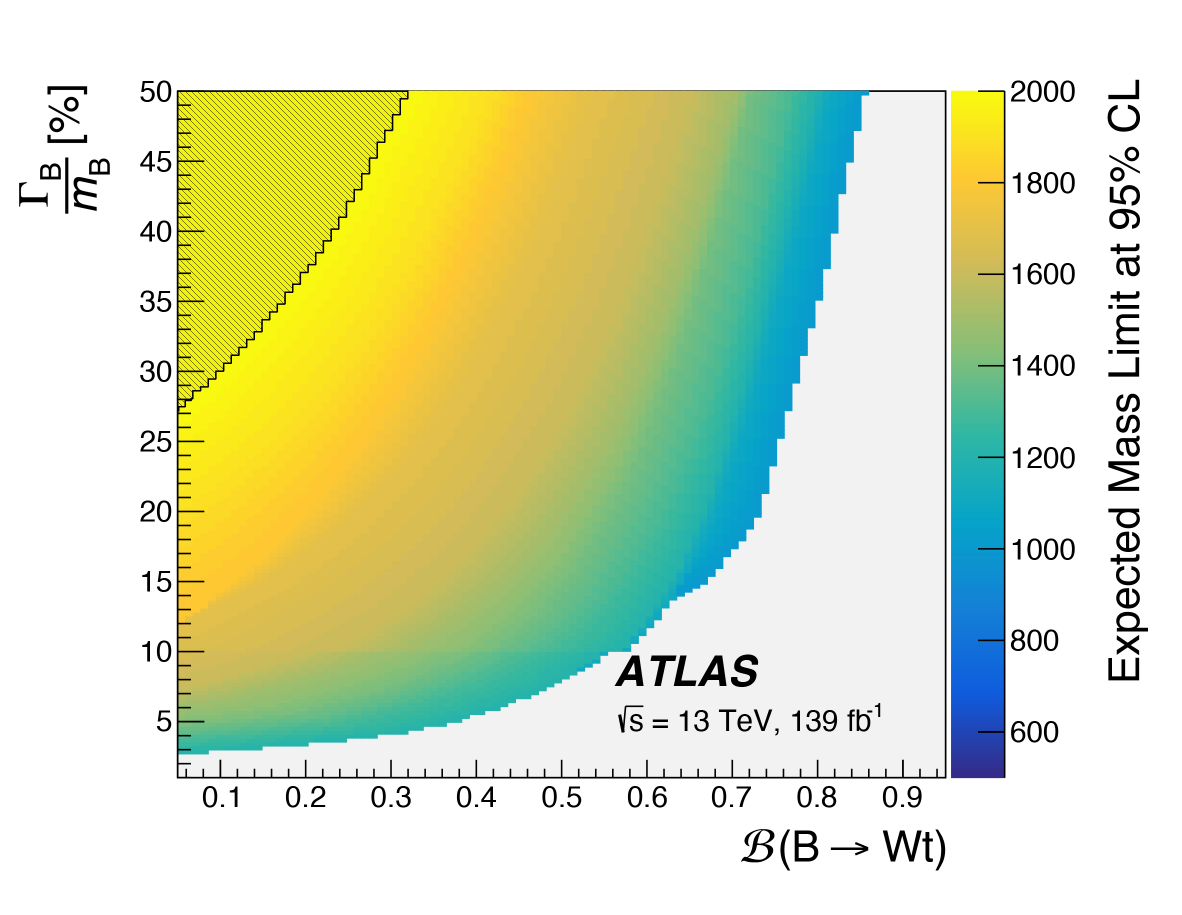}}\label{fig:newstyle_exp}}
 
\subfloat[]{{\includegraphics[width=0.76\textwidth]{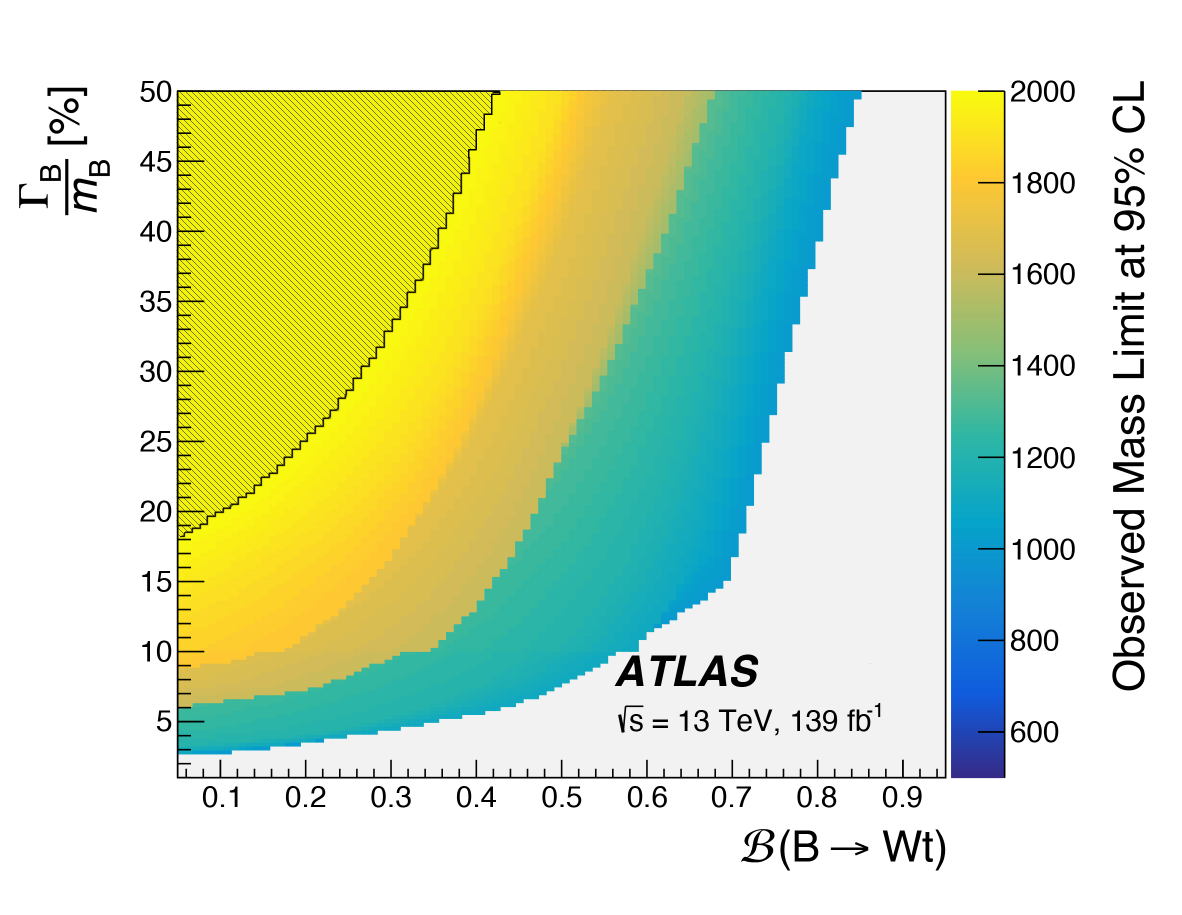}}\label{fig:newstyle_obs}}
\caption{Expected (a) and observed (b) lower limit on the VLB resonance mass given a specific configuration in the phenomenological $(\mathcal{B}(B\rightarrow tW),\ \Gamma_B/m_B)$ phase space. The light grey area corresponds to configurations for which no exclusion is achieved, while the yellow area with the black overlaid pattern corresponds to the phase-space region where every resonance mass scenario considered by the analysis is excluded.}
\label{fig:new_style}
\end{figure}

Figure~\ref{fig:new_style} shows the expected and observed lower limits on the VLB resonance mass in the phenomenological phase space defined above. The constraints on the values of the $\xi$ parameters imply that any specific value of $\mathcal{B}(B\rightarrow tW)$ uniquely determines the two remaining branching fractions $\mathcal{B}(B\rightarrow bZ)$ and $\mathcal{B}(B\rightarrow bH)$.
The previously examined singlet and $(B,Y)$ doublet scenarios can be extracted from Figure~\ref{fig:new_style} by projecting the limits along vertical lines defined by $\mathcal{B}(B\rightarrow tW) = 0.5$ and $\mathcal{B}(B\rightarrow tW) = 0$ respectively. The light grey areas in Figure~\ref{fig:new_style} correspond to configurations for which no VLB masses are excluded, while the yellow areas with the striped pattern overlaid correspond to regions where every theoretical scenario covered by the analysis is excluded.
The discontinuity in the observed mass limits (displayed as a sharp transition between yellow and turquoise in Figure~\ref{fig:newstyle_obs}) is understood to originate from the juxtaposition of the small excess of observed events with $m_B \sim 1.3~\TeV$ and the small deficit for $1.4~\TeV < m_{B} < 2.0~\TeV$.

\clearpage
 
\section{Conclusions}
\label{sec:conclusion}

 
A search is presented for single-production of a vector-like $B$ quark decaying into a Standard Model $b$-quark and Higgs boson, itself decaying into a pair of bottom quarks. The search is carried out in proton--proton collision data with a centre-of-mass energy $\sqrt{s} = 13~\TeV$ collected by the ATLAS experiment at the LHC between 2015 and 2018, corresponding to a total integrated luminosity of $139~\mathrm{fb^{-1}}$.

The search results are interpreted in terms of $95\%$ CL mass-dependent upper limits on the $B$ production cross-section in a number of theoretical scenarios determined by the value of the coupling strength $\kappa$ and the isospin multiplet state. The occurrence of a VLB as an isospin singlet is excluded by this search for values of the main coupling $c_W$ greater than 0.4 for a resonance with $m_B = 1.1~\TeV$, whereas in the $1.3~\TeV < m_{B} < 2.0~\TeV$ range the lowest excluded values of $c_W$ range from 0.5 to 0.6.
 
Owing to the larger single-VLB production cross-section times $B\to bH$ branching fraction, the search sets much stronger limits on a VLB occurring as part of a $(B,Y)$ isospin doublet, with the exclusion limit on the relevant parameter $c_Z$ varying from 0.3 and 0.5 across the investigated resonance mass range $1.0~\TeV < m_{B} < 2.0~\TeV$.
Additionally, the search result is interpreted more generally in terms of upper bounds on the  $B$ production cross-section as a function of the resonance mass and coupling strength $\kappa$, and in terms of excluded resonance mass as a function of the branching fraction into one of the three possible $B$ decay modes and the resonance's total relative width.
 
This search improves on the previously published searches by CMS in the $B\rightarrow bH$ channel, significantly expanding the region of the VLQ theoretical phase space explored and excluded by collider experiments.


\section*{Acknowledgements}


We thank CERN for the very successful operation of the LHC, as well as the
support staff from our institutions without whom ATLAS could not be
operated efficiently.
 
We acknowledge the support of
ANPCyT, Argentina;
YerPhI, Armenia;
ARC, Australia;
BMWFW and FWF, Austria;
ANAS, Azerbaijan;
CNPq and FAPESP, Brazil;
NSERC, NRC and CFI, Canada;
CERN;
ANID, Chile;
CAS, MOST and NSFC, China;
Minciencias, Colombia;
MEYS CR, Czech Republic;
DNRF and DNSRC, Denmark;
IN2P3-CNRS and CEA-DRF/IRFU, France;
SRNSFG, Georgia;
BMBF, HGF and MPG, Germany;
GSRI, Greece;
RGC and Hong Kong SAR, China;
ISF and Benoziyo Center, Israel;
INFN, Italy;
MEXT and JSPS, Japan;
CNRST, Morocco;
NWO, Netherlands;
RCN, Norway;
MEiN, Poland;
FCT, Portugal;
MNE/IFA, Romania;
MESTD, Serbia;
MSSR, Slovakia;
ARRS and MIZ\v{S}, Slovenia;
DSI/NRF, South Africa;
MICINN, Spain;
SRC and Wallenberg Foundation, Sweden;
SERI, SNSF and Cantons of Bern and Geneva, Switzerland;
MOST, Taiwan;
TENMAK, T\"urkiye;
STFC, United Kingdom;
DOE and NSF, United States of America.
In addition, individual groups and members have received support from
BCKDF, CANARIE, Compute Canada and CRC, Canada;
PRIMUS 21/SCI/017 and UNCE SCI/013, Czech Republic;
COST, ERC, ERDF, Horizon 2020, ICSC-NextGenerationEU and Marie Sk{\l}odowska-Curie Actions, European Union;
Investissements d'Avenir Labex, Investissements d'Avenir Idex and ANR, France;
DFG and AvH Foundation, Germany;
Herakleitos, Thales and Aristeia programmes co-financed by EU-ESF and the Greek NSRF, Greece;
BSF-NSF and MINERVA, Israel;
Norwegian Financial Mechanism 2014-2021, Norway;
NCN and NAWA, Poland;
La Caixa Banking Foundation, CERCA Programme Generalitat de Catalunya and PROMETEO and GenT Programmes Generalitat Valenciana, Spain;
G\"{o}ran Gustafssons Stiftelse, Sweden;
The Royal Society and Leverhulme Trust, United Kingdom.
 
The crucial computing support from all WLCG partners is acknowledged gratefully, in particular from CERN, the ATLAS Tier-1 facilities at TRIUMF (Canada), NDGF (Denmark, Norway, Sweden), CC-IN2P3 (France), KIT/GridKA (Germany), INFN-CNAF (Italy), NL-T1 (Netherlands), PIC (Spain), ASGC (Taiwan), RAL (UK) and BNL (USA), the Tier-2 facilities worldwide and large non-WLCG resource providers. Major contributors of computing resources are listed in Ref.~\cite{ATL-SOFT-PUB-2023-001}.


\clearpage

\printbibliography

\clearpage
\input{atlas_authlist}
\end{document}

%% file: atlas_authlist.tex
 
\begin{flushleft}
\hypersetup{urlcolor=black}
{\Large The ATLAS Collaboration}

\bigskip

\AtlasOrcid[0000-0002-6665-4934]{G.~Aad}$^\textrm{\scriptsize 100}$,
\AtlasOrcid[0000-0002-5888-2734]{B.~Abbott}$^\textrm{\scriptsize 117}$,
\AtlasOrcid[0000-0002-7248-3203]{D.C.~Abbott}$^\textrm{\scriptsize 101}$,
\AtlasOrcid[0000-0002-2788-3822]{A.~Abed~Abud}$^\textrm{\scriptsize 34}$,
\AtlasOrcid[0000-0002-1002-1652]{K.~Abeling}$^\textrm{\scriptsize 53}$,
\AtlasOrcid[0000-0002-2987-4006]{D.K.~Abhayasinghe}$^\textrm{\scriptsize 92}$,
\AtlasOrcid[0000-0002-8496-9294]{S.H.~Abidi}$^\textrm{\scriptsize 27}$,
\AtlasOrcid[0000-0001-5329-6640]{H.~Abramowicz}$^\textrm{\scriptsize 148}$,
\AtlasOrcid[0000-0002-1599-2896]{H.~Abreu}$^\textrm{\scriptsize 147}$,
\AtlasOrcid[0000-0003-0403-3697]{Y.~Abulaiti}$^\textrm{\scriptsize 5}$
% ...
\bigskip
\\

% ...
$^{5}$High Energy Physics Division, Argonne National Laboratory, Argonne IL; United States of America.\\
% ...
$^{27}$Physics Department, Brookhaven National Laboratory, Upton NY; United States of America.\\
% ...
$^{34}$CERN, Geneva; Switzerland.\\
% ...
$^{53}$II. Physikalisches Institut, Georg-August-Universit\"{a}t G\"ottingen, G\"ottingen; Germany.\\
% ...
$^{92}$Department of Physics, Royal Holloway University of London, Egham; United Kingdom.\\
% ...
$^{100}$CPPM, Aix-Marseille Universit\'e, CNRS/IN2P3, Marseille; France.\\
$^{101}$Department of Physics, University of Massachusetts, Amherst MA; United States of America.\\
% ...
$^{117}$Homer L. Dodge Department of Physics and Astronomy, University of Oklahoma, Norman OK; United States of America.\\
% ...
$^{147}$Department of Physics, Technion, Israel Institute of Technology, Haifa; Israel.\\
$^{148}$Raymond and Beverly Sackler School of Physics and Astronomy, Tel Aviv University, Tel Aviv; Israel.\\
% ...

\end{flushleft}
